\documentclass[manuscript,screen,review=false]{acmart}
\usepackage{multirow}
\usepackage{tabularx}
  \newcolumntype{L}{>{\raggedright\arraybackslash}X}
\AtBeginDocument{%
  \providecommand\BibTeX{{%
    \normalfont B\kern-0.5em{\scshape i\kern-0.25em b}\kern-0.8em\TeX}}}

\setcopyright{acmcopyright}
\copyrightyear{2021}
\acmYear{2021}
\acmDOI{10.1145/1122445.1122456}

\begin{document}

\title{How many FIDO protocols are needed? Surveying the design, security and market perspectives}

\author{Anna Angelogianni}
\affiliation{%
  \institution{University of Piraeus}
  \streetaddress{80, M. Karaoli \& A. Dimitriou St.}
  \city{Piraeus}
  \country{Greece}}
\email{angelogianni@unipi.gr}

\author{Ilias Politis}
\affiliation{%
  \institution{University of Piraeus}
  \streetaddress{80, M. Karaoli \& A. Dimitriou St.}
  \city{Piraeus}
  \country{Greece}}
\email{ipolitis@ssl-unipi.gr}

\author{Christos Xenakis}
\affiliation{%
  \institution{University of Piraeus}
  \streetaddress{80, M. Karaoli \& A. Dimitriou St.}
  \city{Piraeus}
  \country{Greece}}
\email{xenakis@unipi.gr}

\renewcommand{\shortauthors}{Angelogianni, et al.}

\begin{abstract}
Unequivocally, a single man in possession of a strong password is not enough to solve the issue of security. Studies indicate that passwords have been subjected to various attacks, regardless of the applied protection mechanisms due to the human factor. The keystone for the adoption of more efficient authentication methods by the different markets is the trade-off between security and usability. To bridge the gap between user friendly interfaces and advanced security features, the Fast Identity Online (FIDO) alliance defined a number of authentication protocols. Although FIDO’s biometric based authentication is not a novel concept, still daunts end users and developers, which maybe a contributor factor obstructing FIDO’s complete dominance of the digital authentication market.This paper traces the evolution of FIDO protocols, by identifying the technical characteristics and security requirements of the FIDO protocols throughout the different versions while providing a comprehensive study on the different markets (e.g., digital banking, social networks, e-government, etc.), applicability, easy of use, extensibility and future security considerations. From the analysis we conclude that there is currently no dominant version of a FIDO protocol and more importantly, earlier FIDO protocols are still applicable to emerging vertical services.
\end{abstract}

\maketitle
\section{Introduction}
\label{Intro}
It is a truth universally acknowledged that a single man in possession of a strong password is not enough to solve the issue of security. Literature proves that the classic username and password scheme has been subjected to various attacks, regardless of the applied protection mechanisms. Indeed, either due to limited security awareness from the user's side (i.e., weak or reused passwords etc.) or due to server side vulnerabilities (i.e, vulnerable hash scheme), most data breaches of 2019 have been caused by password-related vulnerabilities. One of the most famous attacks of 2020, affecting the Solar Wings was caused by a weak password in the update server, resulting to huge financial costs apart from the reputational damage \cite{Guardian_SolarWings_Attack}. Phishing attacks are also wider nowadays with an estimated number of 8.3 million attacks for 2019 while during 2020 there was a 667\% increase in phishing scams in only 1 month during the COVID-19 pandemic \cite{enisa_phishing}. This fact dictates that even in cases where the users have chosen relatively strong passwords, it is highly plausible that they might still fall victims of credential theft. To overcome some of the issues caused by passwords, Multi-Factor Authentication (MFA) has been proposed. Nevertheless, according to literature \cite{SIADATI201714} neither MFA has established a secure alternative as short message service (SMS) and voice protocols, that are  typically used to send the One Time Password (OTP) to the user, are not encrypted. To this extend, SIM swapping attacks became popular in latest years, leading to second factor authentication (2FA) bypassing \cite{europol_simswap} \cite{yahoo_finance_simswap}.
\par
Different authentication schemes have been proposed throughout the years however, all of them have failed to replace the existing username and password. There are many reasons behind this failure nevertheless, the most essential is the trade-off between security and usability. Even though modern users are more familiar with the concept of security, therefore desire an elevated level of protection, they are not willing to sacrifice their comfort. This is where FIDO protocol arrives. FIDO comes to combine both security and usability in a proven secure thus fast authentication scheme.
\par
The concept of biometric authentication is not new. Nevertheless, until recent years users were not familiar with these solutions. Furthermore, it is proven by the existing literature that the same notion applies to developers too as due to the lack of experience with such advanced security concepts, they tend to implement them rather unsafely i.e., through faulty implementation of the Android fingerprint API \cite{bianchi2018brokenfingers}. Therefore, a protocol such as the one suggested by the FIDO Alliance, proposes thus validates through its certification process, a common set of requirements \cite{fido_certification}. 
The FIDO protocol suggests a passwordless as well as a 2FA scheme that may use biometric data or security keys instead of passwords. Although the idea of biometric data has been broadly investigated by the existing literature, the FIDO protocol has achieved to encompass it in a modern solution that manages to verify both the user and the server side, preventing phishing among other common attacks \cite{FIDO_research}.
\par
The first FIDO specifications were released during 2014, including i) Universal 2nd Factor (U2F) and ii) Universal Authentication Framework (UAF) definition. Since then, U2F has released two updated versions; v1.1 and v1.2 \cite{fido_u2f_specs} while UAF has also released two updated versions; v1.1 \cite{fido_uaf_specs_v11} and v1.2 \cite{fido_uaf_specs_v12} with the deviation that in UAF v1.2 remains a review draft since 2017. The first working draft from W3C concerning the API for WebAuthn \cite{fido2_specs_webauthn_level1}, was released in 2016 while the FIDO Alliance rolled out the Client To Authenticator Protocol (CTAP2) \cite{fido2_specs_ctap} in 2018. There are working drafts that support an alignment between UAF's v1.2 and FIDO2 metadata nevertheless, the two protocols remain distinct. Even though UAF was released prior to FIDO2, it presents a limited applicability, whereas FIDO2 protocol has been successfully coupled with Internet applications, with seemingly identical security requirements nonetheless. It remains ambiguous though, whether FIDO2 replaces the first FIDO protocols (i.e., UAF and U2F). 
\par
Stemming from this ambiguity and in order to identify the role of the various FIDO protocols within the emerging 6G ecosystem, this paper comprehensively studies and compares the technical and security design and implementation characteristics of the  FIDO protocols, including their message flows, message formats and applications, focusing on their security attributes. This analysis lies the foundations of a in depth critique on the adoption of the different FIDO protocols from different markets and business sectors, the current and future trends and the factors that fuel or prevent their faster adoptions. The paper delves into possible drawbacks or difficulties in the implementation of the FIDO protocols and presents the outlook for future directions that different versions of FIDO should aim for. Focusing on the security aspects of the FIDO implementations, the paper identifies the benefits of FIDO UAF and FIDO2 protocols for application developers and security engineers and underlines the importance of the different FIDO protocols for secure and privacy preserving implementations over the future mobile and IoT networks (i.e., 6G, factories of the future, Industry 4.0/5.0, Industrial IoT, etc.)
\par
The rest of the paper is organised as follows. Sections~\ref{sec:FIDO2_WebAuthn_CTAP2_Overview},~\ref{sec:UAF_Overview} and~\ref{sec:U2FandCTAP1} detail the technical and security characteristics of FIDO2 (WebAuthn \& CTAP2), FIDO UAF and FIDO U2F (CTAP1) protocols, respectively. In Section~\ref{sec:MarketSurvey} a comprehensive survey of the role of FIDO protocols in different markets and industries is presented, while the technical, security and business factors affecting FIDO's adoption is discussed in Section~\ref{sec:critique}. The paper concludes with Section~\ref{sec:conclusion}.

\section{Related Work}
\label{related_work}
Although the research over authentication protocols, multi-factor and passwordless authentication including FIDO, is extensive, there are limited studies focusing on the implementation aspects of FIDO protocols, their applicability in real-world scenarios and the differences they reveal when integrated in services across different verticals (i.e., digital banking, e-governments, mobile applications, etc.). FIDO protocols are present in almost every modern research work that deals with secure and user friendly authentication.
\par
In \cite{barbosaprovable} and \cite{10.1145/3190619.3190640} the overall security guarantees provided by FIDO2/WebAuthn protocol are analyzed, exploring the cryptographic dimension.
Usability studies for FIDO2 passwordless scheme  have been conducted in \cite{FIDO2Kingslayer} and \cite{255646} including real participants. The results of \cite{FIDO2Kingslayer} demonstrated that most users evaluated passwordless authentication as a both usable and secure alternative to passwords nevertheless the participants of this study were particularly concerned about the case where their security key gets lost, stolen or simply damaged. The authors in \cite{255646} further deduced that despite the advanced level of security and usability provided by key-based login, several participants switched back to password manager solutions, built-in their browsers, which were faster. In addition, not all browsers and operating systems supported WebAuthn in the time of the conducted studies therefore some participants could not truly experience passwordless authentication.
\par
The first versions of FIDO protocols have been also widely studied throughout the years. 
In \cite{fengformal} and \cite{10.1007/978-3-319-67639-5_11} an analysis of the UAF protocol from a cryptographic and security perspective is respectively provided. Vulnerabilities and attacks against UAF have been researched in \cite{hu2016security}, \cite{10.1007/978-3-319-67639-5_11},  \cite{fengformal} and \cite{li2020authenticator}. In \cite{Of_two_minds_with_2factor_238317},  \cite{10.1007/978-3-662-58387-6_9} and \cite{238325} usability studies of U2F with external participants are performed.
\par
The research works focusing on the identification and analysis of the differences (protocol design, security provisions, etc.) are limited. The study more closely related to this paper is presented in \cite{10.1145/3319535.3363283}, where the authors analyse the pitfalls of FIDO2/WebAuthn protocol for developers, indicating that lack of sufficient understanding of the protocol,  insecure or incomplete of libraries and privacy concerns are pivotal issues for the community. Nevertheless, the aforementioned study focuses solely on FIDO2/WebAuthn while, this work tries to understand the reasons that led to multiple FIDO protocols and the differences between these versions. The hypothesis this work is based on is that none of the FIDO protocols have been depreciated, in light of the newer releases, yet FIDO2/WebAuthn appears as the most popular.

\section{FIDO2/WebAuthn and CTAP2 Overview}
\label{sec:FIDO2_WebAuthn_CTAP2_Overview}
FIDO2 comprises of the latest addition in the FIDO group of protocols. Although it resembles the first versions of the FIDO protocols (UAF and U2F), FIDO2 is devoted to web browsers that act as FIDO clients. W3C in collaboration with the FIDO Alliance released the WebAuthn API \cite{w3c_webauthn_api}, which is addressed to web app, user agent and authenticator developers intending to support FIDO2 passwordless authentication. Authenticators could be i) \textit{bound} to the device or ii) \textit{external} (i.e., security keys). In order to cover the communication between the FIDO client and the external authenticator, the FIDO Alliance released Client To Authenticator Protocol (CTAP2) specifications. 
\par
WebAuthn supports more than one attestation methods to verify the model identity of the  authenticator: 
\begin{itemize}
  \item \textit{Basic Full Attestation}, where authenticators of the same model share a common attestation private key and attestation certificate. This approach conceals user's identity in the event of a certificate revocation due to private key compromise.
  \item \textit{Basic Surrogate}, also referred to as \textit{Self Attestation} where the authenticator does not have an attestation key therefore it uses the same private key as the one generated for the authentication to the relying party. Basic Surrogate does not provide any cryptographic proof of the authenticator’s security characteristics.
  \item \textit{Attestation CA}, where the TPM holds the authenticator-specific "endorsement key" (EK) used to communicate with a trusted third party, the \textit{Attestation} or \textit{Privacy CA}. The Attestation CA issues an Attestation Identity Key (AIK) certificate for each key pair limiting the distribution of the EK to the relying parties and adding a layer of privacy. 
  \item \textit{Anonymisation CA}, where the CA dynamically generates per-credential attestation certificates which however cannot be used to track the user.
  \item \textit{Elliptic Curve Direct Anonymous Attestation (ECDAA)} even though ECDAA was developed by the members of the FIDO Alliance and was extensively mentioned in the FIDO UAF as well as the first version of the WebAuthn specifications due to its privacy benefits, nevertheless it was omitted from the level 2 version of the WebAuthn API since it did not receive the analogous support from web browsers.
  \item \textit{None}, where the relying party indicates that it does not wish to receive attestation information.
\end{itemize}

The overall architecture of WebAuthn is based on the first version of the protocols (i.e., UAF and U2F). On the \textit{relying party’s (RP) side} there is the server that maintains the requested service and handles the authentication procedure. The relying party server may also communicate with the FIDO Alliance Metadata Service (MDS) in order to assess the authenticators \cite{fido_metadata_service}. Nevertheless, this communication is optional and out of scope for WebAuthn. On the \textit{client’s side} there is i) the relying party’s JavaScript application, ii) the browser which acts as the FIDO client (alike the one described in FIDO UAF), iii) the platform (corresponding to UAF's Authenticator Specific Module) and iv) the authenticator. The authenticator may support a variety of user verification methods such as fingerprint, hand-print, voice-print, face-print, eye-print, pattern, passcode or even location. User verification and user presence are treated as two distinct procedures in FIDO. In addition, \textit{silent authenticators}, which neither authenticate the user nor verify the user's presence, are described. A main difference of between WebAuthn and UAF, is that in WebAuthn, the messages exchanged between the browser (FIDO client) and the platform (ASM), are not declared in the specifications. The proposed API provides a level of abstraction. Nevertheless, judging by UAF, the nature of this messages is focused on error codes indicating specific problems from the authenticator's side and FIDO supported version statements. WebAuthn specifications further declare that the discovery of the transports supported by a given authenticator is outside the scope of the protocol for user agents.
\par
The fundamental idea is that the credentials that belong to a specific user are managed by a trusted authenticator, with which the WebAuthn relying party interacts through the client platform. Relying party scripts can (with the user’s consent) request the browser to create a new credential for future authentication to the relying party. Registration and authentication are performed in the authenticator and mediated by the client platform. The client platform does not have access to the credentials, it can only access information regarding the form or type of the objects. Similarly the relying party does not possess the user verification information (i.e., fingerprint) just the signed response produced by the authenticator. The authenticator may also implement a user interface for the management of the credentials (reset password, history, saved passwords, cookies or even credential deletion). WebAuthn does not explicitly define the deregistration flow.

\subsection{Authenticator Registration (Credentials Create)}
\label{subsec:FIDO2Registration}
The scope of registration is to enroll the authenticator to the relying party and verify the validity of its attestation private key as well as its security characteristics.
\begin{enumerate}
  \item \textbf{Public key Credential Creation - navigator.credentials.create: }The user navigates to the website that offers the requested service in a browser and chooses to “register a security key”. The registration operation is initiated by the relying party which calls the \texttt{navigator.credentials.create} function to request the creation of a new public key credential source. The \texttt{navigator.credentials.create} includes 
  \begin{itemize}
      \item the \textit{origin} of the relying party (e.g., domain, ip, host)
      \item the \textit{sameOriginWithAncestors} which is set to "true" if the execution environment is the same with the caller's and "false" for cross-origin
      \item the \textit{options}. The options object for the creation of the public key credentials include data about i) the RP entity information (\textit{RPid} and \textit{name}), ii) the \textit{user id}, and the \textit{displayed name}, iii) the random \textit{challenge}, iv) the public key credential parameters which denote the \textit{type} (currently only one credential type is defined, and that is "public-key") and the \textit{algorithm} (i.e., ECDSA, RSA) that must be used for the credentials, v) the \textit{timeout} denoted the time that the rp awaits for a response though this can be overwritten by the client, vi) the \textit{exclude credentials} which if present, then the \textit{type} (“public-key”), the \textit{id} plus the \textit{transport}, meaning the communication protocol with the authenticator (e.g., USB, NFC, BLE or internal authenticator) of the excluded credentials are defined, vii) the \textit{authenticator selection criteria}, indicating the relying party's requirements on the authenticator. More specifically the \textit{authenticantor attachment} which indicates if the relying party prefers the authenticator to be internal or cross-platform, the \textit{resident key} if required by the relying party, if the relying party requires to firstly identify the user to provide the user’s credential IDs during authentication, or if it requires \textit{user verification} viii) the \textit{attestation preference} regarding the authenticator's attestation statements (i.e., “none”, "direct” as generated from the authenticator, or “indirect”  to protect user's privacy or "enterprise" for controlled deployments) and lastly, ix) the \textit{extensions}. 
  \end{itemize}
  The user agent checks on the presence and validity of the received information and creates the \textit{ClientData instance} which includes i) the \textit{type} which is “webauthn.create", ii) the RP’s \textit{challenge}, iii) the \textit{origin} and iv) the \textit{token binding} which is optional and indicates the protocol used for the communication of the user agent with the relying party. If the client supports token binding, then its status is “present” and the id is set to a valid string. Up to this day, no major browser has implemented the token binding feature though \cite{hacking_multifactor_authentication}. The client data is serialized in JSON format and hashed in ClientDataHash using SHA-256. Afterwards, the user agent starts the \textit{timer} from the timeout parameter, locates the available authenticators and checks which ones match the \textit{authenticatorSelection}. The user agent selects the first authenticator in the list that matches the criteria and it is not included in the \textit{excludeCredentialDescriptionList}. The authenticator will be appended in the \textit{issuedRequests} until the process is finished.
  \item \textbf{Authenticator Make Credential: }The authenticator make credential operation is initiated by the FIDO client which sends to the authenticator the \textit{ClientDataHash} and all the information that it received from the relying party except for the timeout and the authenticator selection criteria as this information is managed exclusively by the client. More specifically, the  following parameters are sent for the client to the authenticator: 
  \begin{itemize}
      \item the \textit{ClientDataHash}
      \item the \textit{id} of the relying party (RPEntity)
      \item the \textit{user entity}, meaning the \textit{id} and the \textit{display name} of the user
      \item whether the client satisfies the requirement of the relying party on a \textit{resident key}
      \item the requirement on \textit{user presence}. Currently not present in the latest version of CTAP2.
      \item whether the client satisfies the requirement of the relying party regarding the \textit{user verification} (i.e., required, preferred or discouraged)
      \item the \textit{public key credential types} and the \textit{algorithms} as requested by the relying party
      \item the \textit{exclude credential descriptor list} as provided by the relying party and   
      \item the \textit{extensions} created by the client based on the extensions requested by the relying party.
      \item lately in the Level 2 version of the WebAuthn API, the \textit{Enterprise Attestation Possible} was added to cover individually-identifying attestations
  \end{itemize}
 In CTAP2 protocol i) \textit{pinAuth} and ii) \textit{pinProtocol} are included. The authenticator confirms the validity of the received information and prompts the user to give his/her consent displaying in parallel the id and the name of the relying party as well as the name and the display name of the user’s account. 
  \item \textbf{Attestation Object: }
  Once the user’s consent is acquired, either by verifying the user or by testing his/her presence, the authenticator creates the \textit{public key pair} (public and private) and the \textit{user handle}, which is a unique id specified by the relying party per user account and facilitates the management of the credentials. The authenticator will initialise the \textit{signature counter} and replies to the client by sending the \textit{attestation object} which contains: 
  \begin{itemize}
      \item the \textit{authenticator data}. The authenticatior data more specifically, contain i) the \textit{hashed RPid}, ii) the \textit{flags} which signify the user presence, verification and if the attested data or extensions are included, iii) the \textit{signature counter}, iv) the \textit{extensions} and v) the \textit{attested credential}. The attested credential refer to the \textit{AAGUID}, the \textit{credential id} and its \textit{credential's length} as well as the \textit{credential's public key}.
      \item the \textit{attestation statement format}
  \end{itemize}
  \par
  \item \textbf{Authenticator Attestation Response: } The FIDO Client after receiving the response from the authenticator, constructs the \textit{clientDataJSON} object which includes: i)  the \textit{type} of the request (i.e., webautn.create), ii) the \textit{challenge}, iii) the \textit{origin} of the relying party and iv) the \textit{tokenBinding} between the origin and the client. The FIDO Client will forward to the relying party the following data:
  \begin{itemize}
      \item the \textit{attestation object} as received from the authenticator
      \item the \textit{clientDataJSON} object
      \item the \textit{transport} protocol used by the authenticator (i.e., NFC, BLE etc.)
      \item the client \textit{extensions}
  \end{itemize}
The RP, after receiving the publicKeyCred object, will verify the clientData as well as the hash of the clientDataJson and decode the attestation Object to verify its validity. The flow is illustrated in Fig.~\ref{fig:2_Registration_flow}.
\end{enumerate}
\begin{figure}[t]
\centering
\includegraphics[width=0.85\textwidth]{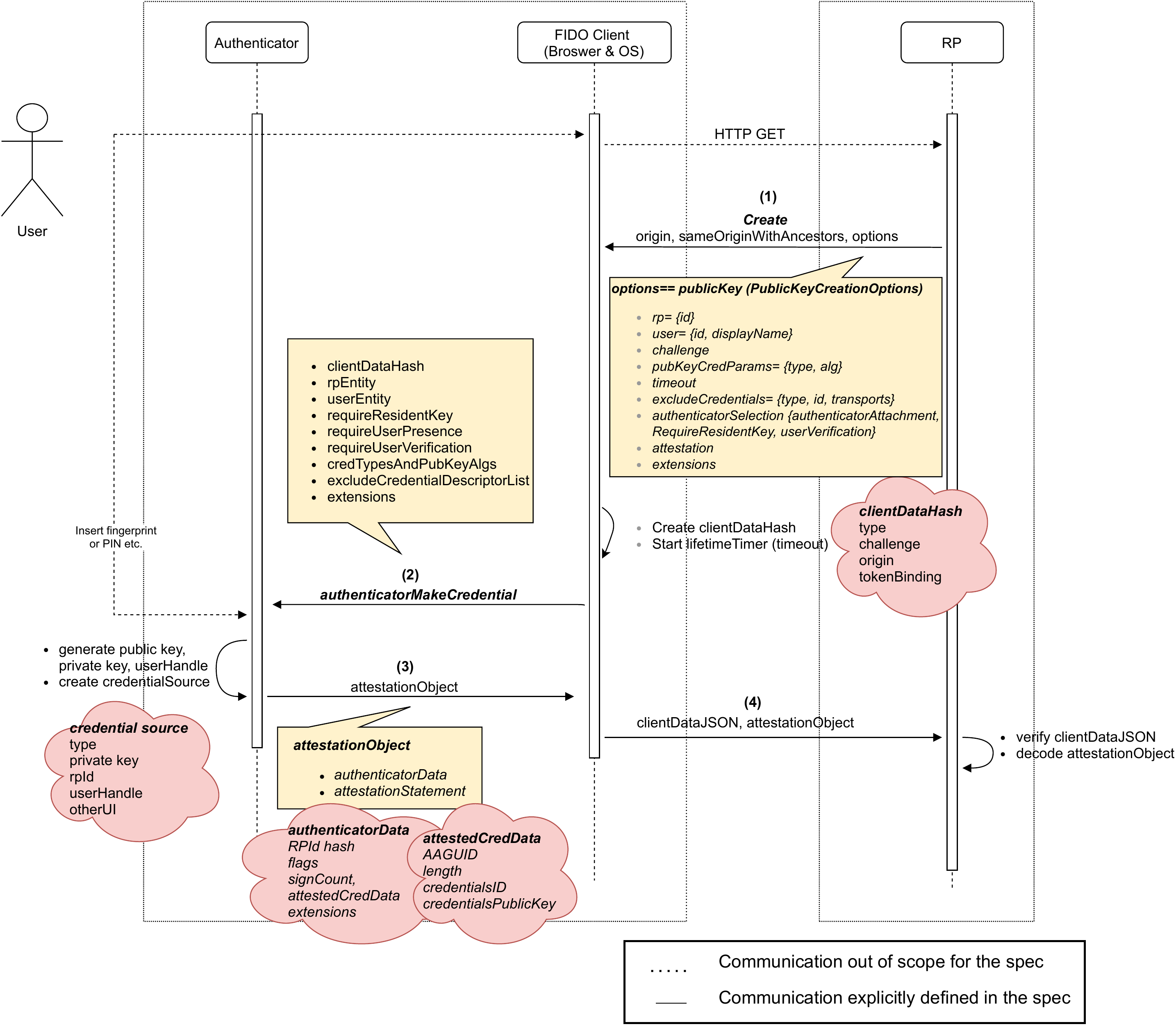}
\caption{FIDO2 Authenticator Registration Message Flow}
\label{fig:2_Registration_flow}
\end{figure}

\subsection{User Authentication (Credentials Get)}
\label{subsec:FIDO2Authentication}
\begin{figure}[t]
\centering
\includegraphics[width=0.9\textwidth]{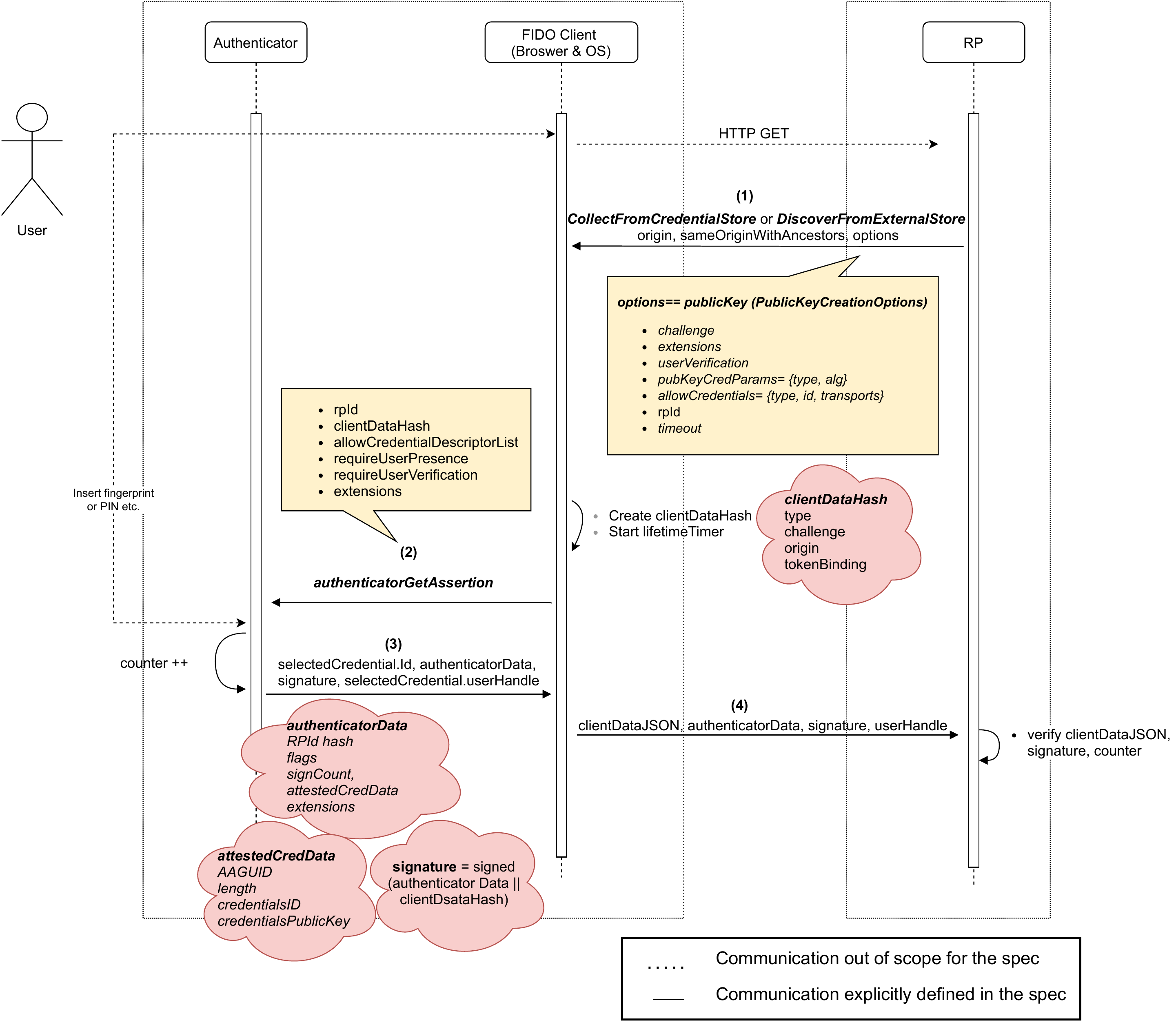}
\caption{FIDO2 User Authentication Flow}
\label{fig:2_UserAuthentication_flow}
\end{figure}

\begin{enumerate}
\item \textbf{Make an Assertion - navigator.credentials.get: }
There are 2 functions that can be used in order to retrieve previously registered credentials: [CollectFromCredentialStore] or [DiscoverFromExternalStore] depending on the authenticator, whether it is external or not. In both cases, the functions contain the same parameters (origin, options, sameOriginWithAncestors) the difference is that in the case of the external authenticator, both functions will be needed when the credentials will not be found within the device. The message sent from the FIDO Server to the FIDO Client includes 
\begin{itemize}
    \item the \textit{origin} (i.e., domain). The options for the creation of the public key credentials include data about i) the RP, such as the \textit{name}, ii) the user such as the \textit{user’s id}, its \textit{name} and \textit{displayed name} iii) the random \textit{challenge}, iv) the public key credential parameters which denote the \textit{type} (“public-key”) and the \textit{algorithm} (i.e., ECDSA) that must be used for the credentials, v) the \textit{timeout}, vi) the \textit{exclude credentials} which if present, they define the type (“public-key”), the \textit{id} plus the \textit{transport}, meaning the communication protocol with the authenticator (i.e., usb, nfc, Bluetooth or internal authenticator), vii) the \textit{authenticator selection criteria}, indicating the requirements on the authenticator type for example if it’s an internal authenticator or if the client device requires a wrapped resident key in order to offload the authenticator or if it requires user verification, viii) the \textit{attestation preference} on the authenticator attestation statements which can be set to “none”, “indirect” for privacy or “direct” and lastly, ix) the \textit{extensions}. 
    \item the \textit{sameOriginWithAncestors} which indicates the execution environment (i.e., browsing context, navigation request)
    \item the \textit{options}
\end{itemize}
\par

\par
The User Agent will check the presence and validity of the values received and create the ClientDataJSON (ClientData in JSON format) instance will be later hashed.
Afterwards, the User Agent will start the lifitimeTimer from the timeout parameter, find the available authenticators and check their options to determine whether any public key credentials are bound to this authenticator. This validation is achieved using the RPId and appending it to the authenticator's issuedRequests.
\textbf{Authenticator Get Assertion: } The FIDO Client will send the autthenticatorMakeCredential to the authenticator which, includes:
\begin{itemize}
    \item the \textit{hashed clientData},
    \item the \textit{id} of the relying party, 
    \item the \textit{user entity},
    \item the requirement of the device on the \textit{user presence},
    \item the requirement on \textit{user verification}, 
    \item the public key credential \textit{types} and the \textit{algorithms}, 
    \item the \textit{exclude credential descriptor list} and 
    \item the \textit{extensions} of the client
\end{itemize}
The authenticator will prompt the user to enter his/her credentials (i.e., fingerprint), check the received parameters and if no error code is reported, it will increase the counter.
\item \textbf{Assertion : }The authenticator will send its response to the client which encapsulates: 
\begin{itemize}
    \item the \textit{selected credential id}, 
    \item the \textit{authenticator data}
    \item the \textit{signature} and 
    \item the \textit{user handle} of the selected credential
\end{itemize}

\par
The \textit{authenticator data} more specifically, contain the hashed RPid, the flags, the signature counter, the extensions and the attested credential. The attested credential refer to the AAGUID, the the credential id and its length as well as the credential's public key. The signature on the other hand refers to the signed disjunction of the authenticator data with the client data hash.
 \item \textbf{Assertion Response: }
The FIDO Client forwards the information received from the authenticator adding the 
\begin{itemize}
    \item the \textit{clientData}
    \item the client \textit{extensions}
\end{itemize}
The RP, after receiving the response will verify the clientData, the hashed clientData, the challenge, the origin, the type of the request, the token binding, the RPId, the user verification and user presence flags as well as the counter and if the received data are indeed valid, then the authentication ceremony will continue else it will fail. The flow is demonstrated in Fig. \ref{fig:2_UserAuthentication_flow}. 
\end{enumerate}

\subsection{Advanced Security}
\label{subsec:Advanced_sec_FIDO2}
The request from the relying party to the client contains information regarding the relying party, whether or not it is a cross origin object as well as information regarding the user. This fact suggests that the credential is issued for a specific username and origin (i.e., relying party) therefore, phishing attacks will be rejected by the authenticator. Similarly, the randomly generated challenge protects from replay attacks. It is of utmost importance that the relying party has the option to choose the characteristics of the authenticator (\textit{authenticatorSelection}) hence, it may exclude authenticators that do not match the chosen criteria or even revoke authenticators where vulnerabilities have been discovered. To this extend a certain security background is required by the developing and maintenance team.
During step \textbf{(2)}, as demonstrated in the Fig.~\ref{fig:2_Registration_flow} and Fig.\ref{fig:2_UserAuthentication_flow}, the FIDO Client forwards most of the information received from the relying party to the authenticator with the exception of the clientData, which is send as a hashed object. The signature counter provides both synchronisation and protection from replicated authenticators. The clientDataHash as well as the counter provide  protection in the case of a cloned authenticator or credential disclosure. 
Additionally, as the RPid is hashed after exiting the authenticator, and the client forwards it within the attestationObject, the relying party may also verify the validity of the client, indicating round-trip integrity. Lastly, the relying party will assure that the authenticator matches indeed the desired criteria by inspecting the (\textit{AAGUID}).
\par 
During the authentication, the credential id and user handle is also forwarded from the authenticator to the relying party in order to verify that these values match the stored ones (send during the registration of the authenticator). Moreover, the authenticator sends a signed object containing the authenticatorData and the hashed clientData, while the client also sends the clientData to the relying party. As a result, the relying party may compare the received values with the expected values and even identify whether a modification took place either in the authenticator-level or in the FIDO client-level, achieving round-trip integrity.

\section{FIDO Universal Authentication Framework (UAF) Overview}
\label{sec:UAF_Overview}
In UAF the supported authenticator attestation methods methods, include, the Basic Full Attestation, the Basic Surrogate and the Elliptic Curve Direct Anonymous Attestation (ECDAA). Similar to FIDO2, UAF communication has two sides: i) the client and ii) the server. In FIDO UAF, the client is composed of: i) the RP’s web application (i.e., frontend), ii) the FIDO Client, iii) the ASM (Authenticator Specific Module) and iv) the authenticator. In comparison with the FIDO2 where the ASM is omitted in view of a more granular approach, UAF defines it as a distinct component. The server on the other hand contains the relying party’s web application server (backend) and the FIDO Server. The client side is responsible for the generation, storage and handling of the credentials while the server’s obligation contains the generation of the challenge as well as the validation and storage of the information received from the client. 
\par
The FIDO Client implements the client’s side FIDO UAF protocol and acts as a midpoint between the server and the authenticator. The ASM is the translator between the FIDO Client and the authenticator that interprets the FIDO UAF messages to authenticator commands. One of the most essential actors in FIDO UAF is the authenticator. It can be an \textit{embedded, bound} or an \textit{external device (roaming authenticator)} while it may support first or/and second factor authentication. The difference between an external and a roaming authenticator is that while in the external the keys and the counter will be stored within the authenticator, in the roaming, the keys will be stored in the ASM.  The relying party contains the web server with the service that the user has requested access to and the FIDO Server which handles the authentication procedure. 
\par
FIDO UAF defines four processes i) authenticator registration, ii) user authentication, iii) transaction confirmation and iv) authenticator deregistration.

\subsection{Authenticator Registration in FIDO UAF}
\label{subsec:UAF_AuthenticatorRegistration}
If the user is already registered to the relying party’s web server he/she will enter its credentials and they are verified then the latter will send a \textit{header} to the FIDO Server which includes information on i) the \textit{UAF version} (upv), ii) the \textit{operation} (op) which in this case is registration, iii) the \textit{ID of the application} (appID) and iv) the \textit{serverData} which is a session identifier created by the relying party (i.e., expiration times for the registration session). 
\begin{figure}[t]
\centering\includegraphics[width=0.95\textwidth]{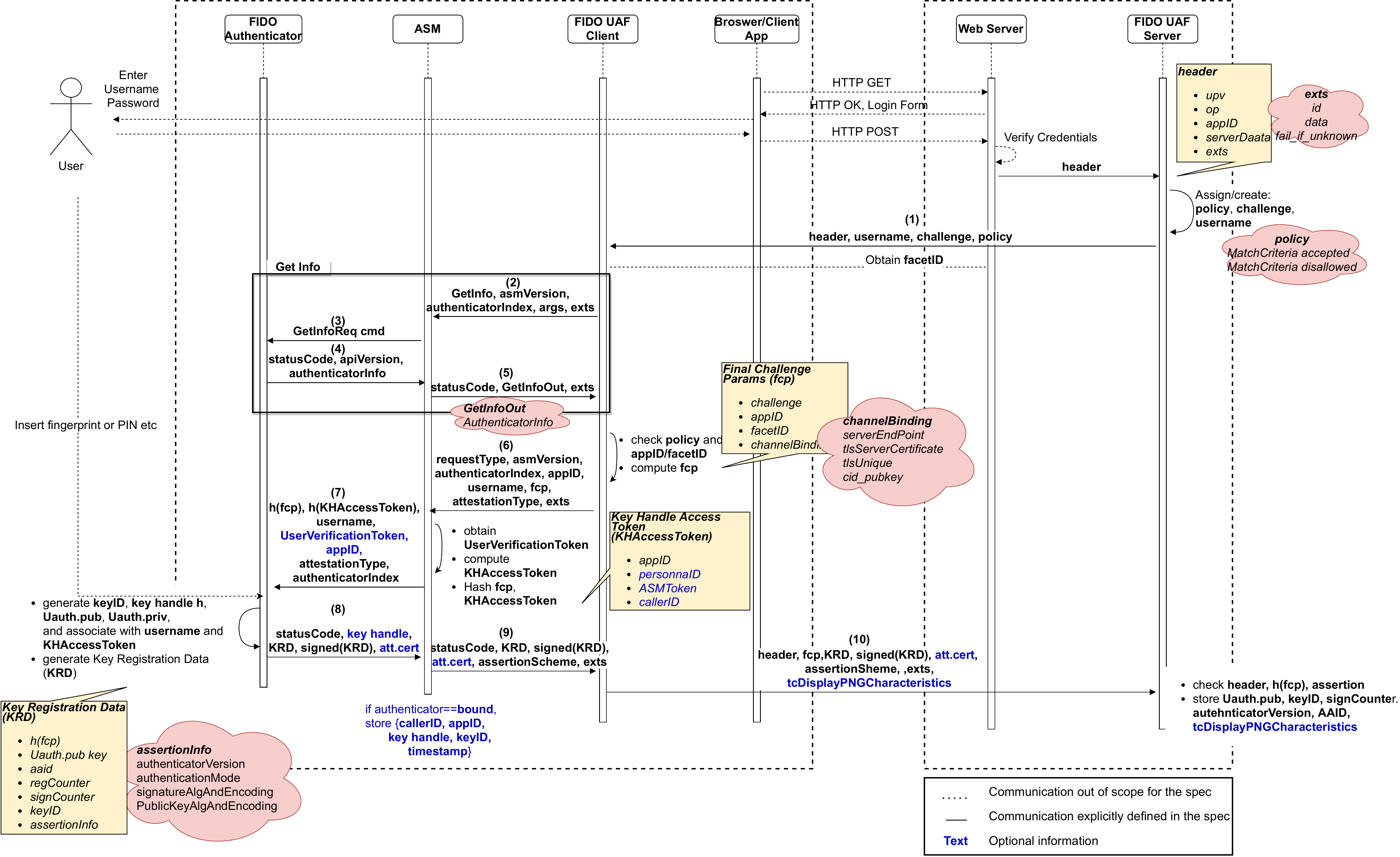}
\caption{FIDO UAF Authenticator Registration Message Flow}
\label{fig:UAF_Registration_flow}
\end{figure}

\begin{enumerate}
  \item \textbf{Registration Request: }After receiving the header from the server of the relying party, the FIDO Server will create the Registration Request message that it will be forwarded to the FIDO Client. The Registration Request encloses: 
  \begin{itemize}
      \item the \textit{header} as sent from the web server of the relying party
      \item the \textit{username} (used to distinguish the different accounts in the event that the user has more than one accounts in the same relying party)
      \item a randomly generated \textit{challenge}
      \item the \textit{policy} which contains both \textit{allowed} and \textit{disallowed} criteria for the authenticators, ordered by highest priority. Among the allowed criteria i) \textit{authenticator attestation id (\textit{aaid})} which refers to the specific authenticator model, the ii) \textit{vendorID} which refers to the authenticator's vendor, the iii) \textit{key ID} (unique for each \textit{aaid}, which is null at this point, since it refers to the user authentication private key which has not been produced yet, the method used for iv) \textit{user verification} (i.e., fingerprint, PIN, etc.),  v) the mechanisms used for \textit{key protection} (i.e., Trusted Execution Environment, Secure Element, etc.) or vi) the method used for the \textit{matcher protection} method which matches the user inserted verification method to the previously stored, the vii) \textit{attachmentHint} indicating the communication protocol between the authenticator and the user device (i.e., internal, Bluetooth, NFC, etc.), the viii) \textit{tcDisplay} designating whether the display is secure therefore it can support Transaction Confirmation, the ix) supported \textit{authentication algorithms} (i.e., ECDSA), the x) supported \textit{assertion scheme} (i.e., KRD for the authenticator registration and signed data for the authentication map to UAFV1TLV), the xi) supported \textit{attestation types} (i.e., Basic Full), the xii) \textit{authenticator version} and the xiii) \textit{extensions}. While, \textit{keyID} is unique per \textit{aaid} but not per user account. This is the reason why UAF includes the \textit{personna} notion to cover different user accounts on the same relying party.
\end{itemize}
  
  \item \textbf{Get Info ASM Request: }The FIDO Client checks the policy and communicate with all available authenticators to find the one that better matches the policy. In order to achieve this, the FIDO Client first sends an ASM Request to the ASM which, contains the \textit{type} of the request (in this case it is a GetInfo request), the \textit{ASM version} indicating the supported FIDO version (i.e., 1.2) and the \textit{extensions} (exts).
 Normally ASM Requests also include the authenticator index and the arguments but in this execution point, these variables are null. 
  \item \textbf{Get Info Command: }The ASM sends the GetInfo Command to the authenticator and the latter responds to the ASM with the Get Info Command Response which contains 
  \begin{itemize}
      \item the \textit{status code} indicating a specific code in case of failure
      \item the \textit{API version} which verifies the supported api version of the authenticator and
      \item the \textit{authenticatorInfo} which includes information regarding i) the \textit{authenticator index} enlisting all available authenticators discovered, ii) the \textit{aaid} and iii) the \textit{authenticator metadata}. The authenticator metadata encapsulates i) the \textit{authenticator type} indicating whether the authenticator is bound or external, first of second factor, ii) the \textit{maximum key handles} that can be processed in a single command, iii) the \textit{user verification} method (i.e., fingerprint), the iv) \textit{key protection} method, v) the \textit{matcher protection} method, vi) the \textit{tcDisplay} designating whether the display is secure therefore it can support Transaction Confirmation, vii) the supported \textit{authentication algorithm}, viii) the supported \textit{assertion scheme} (i.e., KRD for versions 1.0 and 1.1. or signed data available for v1.2) and ix) \textit{attestation type} (i.e., Basic Full), optionally the authenticator may return information on the \textit{supported extensions} and the \textit{tcDisplay type} and \textit{characteristics}. 
  \end{itemize}
  \item \textbf{Get Info ASM Response: }The ASM sends the ASM Response to the FIDO Client containing, the \textit{status code}, the \textit{responseData} (of GetInfoOut type in this case), which encloses the \textit{authenticatorInfo} as send from the authenticator in step (3), showcased in Fig.~ \ref{fig:UAF_Registration_flow} and the \textit{extensions}.
The FIDO Client filters the accepted authenticators according to the policy that where retrieved through the GetInfo request and it verifies that \textit{facetID} of the application is authorized for the \textit{appID}. The facetID could be cached in the FIDO Client’s memory or it can be retrieved from the FIDO Server. Afterwards the FIDO Client computes the \textit{final challenge parameters} (fcp) which include: i) the \textit{challenge}, ii) the \textit{appID}, iii) the \textit{facedID}, and iv) the \textit{channelBinding}, which contains TLS information concerning the communication of the FIDO client with the FIDO server such as the \textit{hashed TLS server certificate}, the \textit{TLS server certificate} and the \textit{channel's public key} (cid\_pubkey).

In UAF v1.2 \textit{ClientData} has been proposed as an alternative to the \textit{fcp} for platforms that support WebAuthn and CTAP2. ClientData includes i) the \textit{challenge}, ii) the \textit{origin} mapped to UAF's facetID, iii) \textit{hashAlgorithm}, iv) \textit{tokenBinding} mapped to UAF's cid\_pubkey of the channelBinding, and v) the \textit{extensions}
  \item \textbf{Register ASM Request: }The FIDO Client sends to the ASM, an ASM Request indicating 
  \begin{itemize}
      \item the \textit{type} of the request, which in this case is “Register”
      \item the supported \textit{ASM version} 
      \item the \textit{index} of the chosen authenticator (authenticatorIndex), which refers to all discovered authenticators
      \item the \textit{args}, which in this case is set to RegisterIn and contains information such as i) the \textit{appID}, ii) the \textit{username}, iii) the \textit{fcp}, iv) the \textit{facetID}, v) the \textit{channel binding data} containing information regarding the TLS ChannelID (cid\_pubkey) and certificate and vi) the \textit{extensions}.
  \end{itemize}
  The ASM locates the authenticator, by using the authenticatorIndex parameter, checks if the user is already enrolled and requests either for user verification (waiting to receive a \textit{UserVerificationToken}) or user enrollment. Next, it hashes the \textit{fcp} and computes and hash the \textit{KHAccessToken} which includes i) the \textit{appID} and if the authenticator is bound ii) the \textit{personnaID} (obtained by the operating system for each user account), iii) the \textit{ASMToken}, which is a random and unique ID generated by the ASM and iii) the \textit{CallerID} which specifies the platform (i.e., IOS).
  \item \textbf{Register Command: }Subsequently, the ASM sends the Register Command to the authenticator which includes: i)the \textit{authenticatorIndex}, ii)the \textit{appID}, iii) the hashed \textit{fcp}, iv)the \textit{username}, v) the \textit{attestationType}, vi) the hashed \textit{KHAccessToken} and vii) the \textit{UserVerifyToken}.
  The authenticator  generates a new \textit{key pair}, the \textit{keyHandle} and the \textit{keyID} and then creates the \textit{key registration data} (KRD) structure which contains i) the \textit{aaid}, ii) the \textit{assertionInfo}, iii) the hashed \textit{fcp}, iv) the \textit{keyID}, v) the \textit{signCounter}, vi) the \textit{RegCounter} and vii) the \textit{UAuth.pub key}. 
  Successively, the authenticator sends the Register Command Response to the ASM which contains the \textit{statusCode}, the \textit{KRD} for UAF v1.0 and 1.1, the \textit{signed KRD} (with the UAuth.priv) for UAF v1.0 and 1.1, the \textit{attestation certificate} (for Basic Full attestation) and the \textit{keyHandle}.
 
  \item \textbf{Register ASM Response: }The ASM will store the callerID, the appID, the keyHandle, the keyID, the currentTimestamp if the authenticator is bound and create the ASM Response comprising of 
  \begin{itemize}
      \item the \textit{statusCode} 
      \item the \textit{responseData} which in this case is set to RegisterOut containing the \textit{assertion} (signifying the \textit{KRD}, the \textit{signed KRD} and the \textit{attestation certificate} if the attestation scheme is Basic Full) as well as the \textit{assertionScheme}. 
  \end{itemize}

  \item \textbf{Registration Response: }The UAF Client will create and send to the FIDO Server the Registration Response holding 
  \begin{itemize}
      \item  the \textit{header} 
      \item the \textit{fcp} or \textit{ClientData} introduced in UAF v1.2
      \item the \textit{assertionScheme}
      \item the \textit{KRD} for UAF v1.0 and 1.1
      \item the \textit{signed KRD} for UAF v1.0 and 1.1 
      \item the \textit{attestation certificate} (\textit{if applicable})
  \end{itemize}
 The FIDO Server will check all components of the Registration Response message and afterwards it will hash the fcp and to if it matches with the hashed fcp included in the KRD. Then, it will store the UAuth.pub, keyID, signCounter, authenticatorVersion and the \textit{aaid}. The FIDO Server may return the result to the web server in order to let it inform the user. The flow is demonstrated in Fig.~\ref{fig:UAF_Registration_flow}.
\end{enumerate}

\subsection{User Authentication in FIDO UAF}
\label{subsec:UAF_Authentication}
As described in the authenticator registration, the process is initiated by the user that wants to access a service and the web server that contains it prompts the user for authentication by forwarding the \textit{header} to the FIDO UAF Server similarly to the Registration procedure including i) the supported \textit{UAF version (upv)}, ii) the \textit{operation (op)} which in this case is authentication, iii) the \textit{appID} and the iv) \textit{serverData}, signifying the TLS information. The flow is depicted in Fig. \ref{fig:UAF_Uauth_Transaction_flow}.
\begin{enumerate}
  \item \textbf{Authentication Request: }The FIDO Server generates the Authentication Request which encloses: 
  \begin{itemize}
    \item the \textit{header}
      \item the randomly generated \textit{challenge}
      \item the \textit{policy} which contains the allowed and disallowed criteria for the authenticators, ordered by highest priority, as described in the Registration step \textbf{1)} with the only exception that the \textit{keyID} is not null. The Relying Party forwards the information to the FIDO Client. 
  \end{itemize}
  
  \item \textbf{Get Info and Authenticate ASM Request: }The FIDO Client obtains the \textit{facetID} and check if it is authorized for the specific \textit{appID}. Afterwards it initiates a GetInfo procedure, as described in the Registration (steps \textbf{2,3}), in order to discover all available authenticators and choose the one suitable according to the policy. After choosing the authenticator, the FIDO Client computes the \textit{fcp} as thorougly presented in the step \textbf{4)} of the Registration
 and subsequently sends an ASM Request to the ASM containing:
\begin{itemize}
     \item the \textit{requestType}, which is set to “Authenticate”
     \item the \textit{asmVersion}
     \item the \textit{authenticatorIndex} which defines the index of the selected authenticator 
     \item the \textit{args} which is set to AuthenticateIn and incorporates i) the \textit{appID}, ii) the \textit{keyID} and the iii) the \textit{fcp}, iv) the \textit{facetID}, v) the \textit{channel binding data} and vi) the \textit{extensions}.
 \end{itemize}
The ASM locates the authenticator, using the \textit{authenticatorIndex} parameter, as well as the \textit{keyHandle} associated with the specific \textit{appID} and \textit{keyID}. Next, it requests for user verification \textit{(UserVerificationToken)} and if this step is successfully completed then the ASM hashes the \textit{fcp} and calculated the hash of the \textit{KHAccessToken} which includes the same information as described in Registration's step \textbf{5)}.
  \item \textbf{Sign Command: }The ASM sends the Sign Command to the authenticator which includes:
  \begin{itemize}
  \item the \textit{authenticatorIndex}
      \item the \textit{appID} (optional)
      \item the \textit{hashed fcp}
      \item the \textit{hashed KHAccessToken} 
      \item the \textit{UserVerifyToken} (optional)
      \item the \textit{key handle}
  \end{itemize}
The authenticator verifies the user (using the retrieved UserVerificationToken), locates the \textit{keyHandle} and use it to validate the \textit{KHAccessToken}. If  the user has more than one accounts, he/she will be asked to choose the username. Next the authenticator  prepares the Sign Command Response which will be send to the ASM. The Sign Command Response encompasses:
\begin{itemize}
      \item the status code and either, \\
      \textbf{Option\_1:}
      \item the \textit{username}
      \item  the \textit{key handle} or \\
      \textbf{Option\_2:}
\item data structure which includes: i) the \textit{aaid}, ii) the \textit{assertionInfo} (e.g., the authenticator version, the authentication mode  indicating whether user explicitly verified, the signature and public key algorithm and encoding format), iii) an authenticator \textit{nonce}, iv) the \textit{hashed fcp}, v) the \textit{keyID}, vi) the \textit{signCounter} and 
\item the data structure signed using the \textit{Uauth.priv}. 
  \end{itemize}
  \item \textbf{Authenticate ASM Response: }The ASM creates the ASM Response which contains 
  \begin{itemize}
    \item the \textit{statusCode}
    \item \textit{responseData} meaning the AuthenticateOut structure which includes the \textit{assertion} meaning the data and signed data) for option 2 of Authentication's step \textbf{3)} and the \textit{assertionScheme}
  \end{itemize}
  \item \textbf{Authentication Response: }After receiving this message, the UAF Client will send the Authentication Response to the FIDO UAF Server including the 
  \begin{itemize}
      \item the \textit{header}
      \item the \textit{fcp}
      \item the \textit{assertionScheme}
      \item the \textit{data}
      \item the \textit{signed data}
  \end{itemize}
  The FIDO Server inspects all the components of the Authentication Response message, hashes the \textit{fcp} in order to compare it with the one included in the data structure and checks whether the \textit{UAuth.pub key}, the \textit{signCounter}, the \textit{authenticatorVersion} and the \textit{aaid} match the ones stored during the registration and verifies the signature. Lastly, the FIDO Server returns an allow-access-to-content message to the web server of the relying party.
\end{enumerate}

\subsection{Transaction Confirmation in FIDO UAF}
\label{subsec:UAF_TransactionConfirmation}
\begin{figure}[ht]
\centering\includegraphics[width=0.85\textwidth]{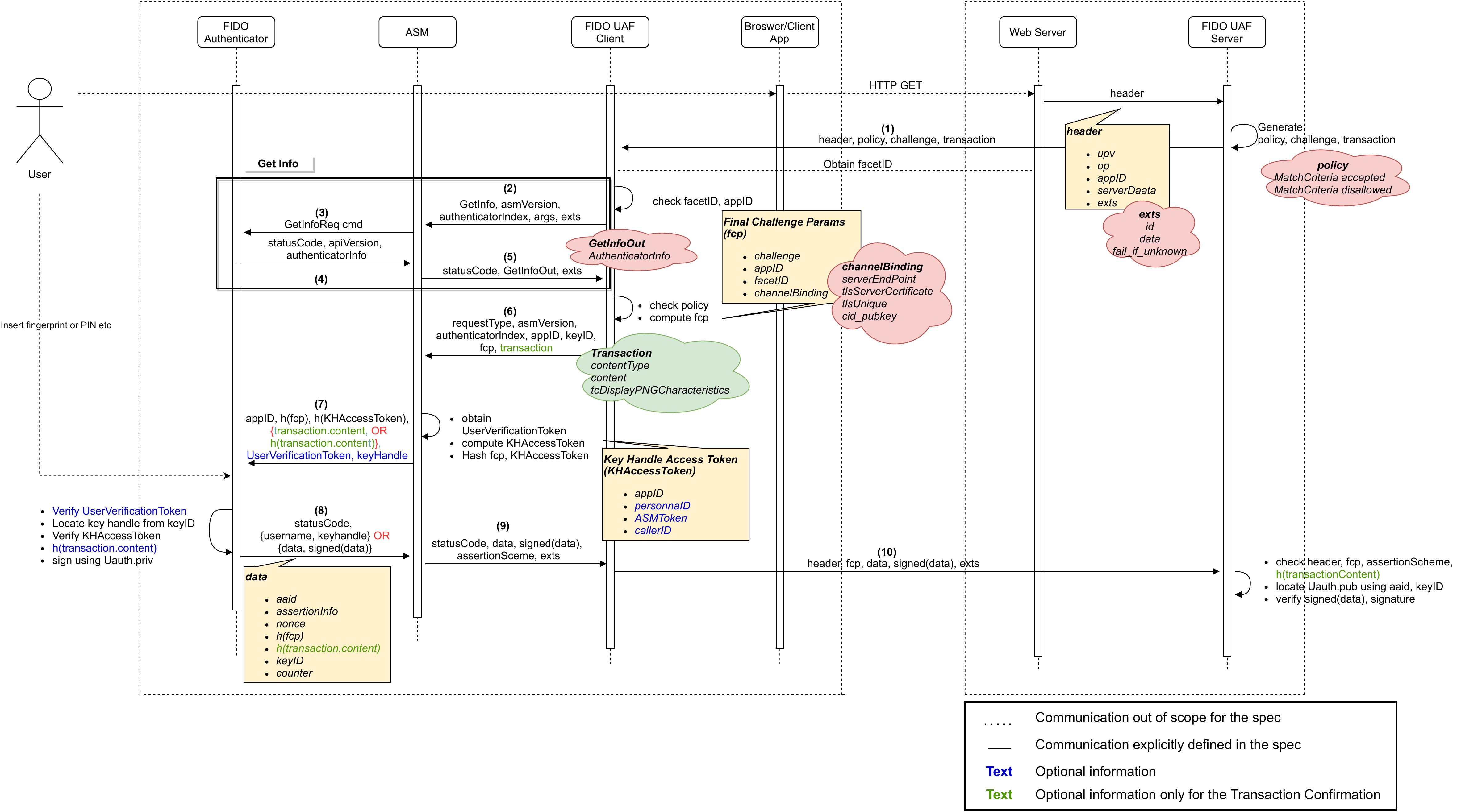}
\caption{FIDO UAF User Authentication and Transaction Confirmation Message Flow}
\label{fig:UAF_Uauth_Transaction_flow}
\end{figure}

Transaction Confirmation has the same message flow as Authentication however some messages consist of additional information that refer mainly to the type of the message, the display characteristics and the transaction content. The flow is illustrated in Fig. \ref{fig:UAF_Uauth_Transaction_flow} in green color.
\begin{enumerate}
    \item \textbf{Authentication Request: }
    In addition to the messages included for the Authentication procedure, in order to provide Transaction Confirmation support, the FIDO Server further includes a \textit{transaction object} which specifies i) the \textit{content type}, ii) the \textit{content} and iii) the \textit{display characteristics} which is sent to the FIDO Client. It should be noted that the \textit{policy} also includes information on the tcDisplay.
    \item \textbf{Get Info and Authenticate ASM Request} The FIDO Client forwards the information related to the transaction to the ASM
    \item \textbf{Sign Command: }The ASM further includes to the Sign Command the \textit{transaction content} or the \textit{hashed transaction content} to the authenticator, which will be verified and signed.
    In the Sign Command Response of \textbf{option 2} the \textit{hashed transaction content} or the \textit{ transaction content} are appended, while the \textit{authentication mode} object of the \textit{assertionInfo} further describes if transaction content has been shown on the display and user confirmed it by explicitly verifying with authenticator. 
    \item \textbf{Authenticate ASM Response: }The signed data containing the information on the transaction content will be later sent back to the FIDO Client.
    \item \textbf{Authentication Response: }The relying party checks the hash of the transaction content for the received and the cached object.
\end{enumerate}

\subsection{Deregistration in FIDO UAF}
\label{subsec:UAF_deregistration}
\begin{figure}[t]
\centering\includegraphics[width=0.7\textwidth]{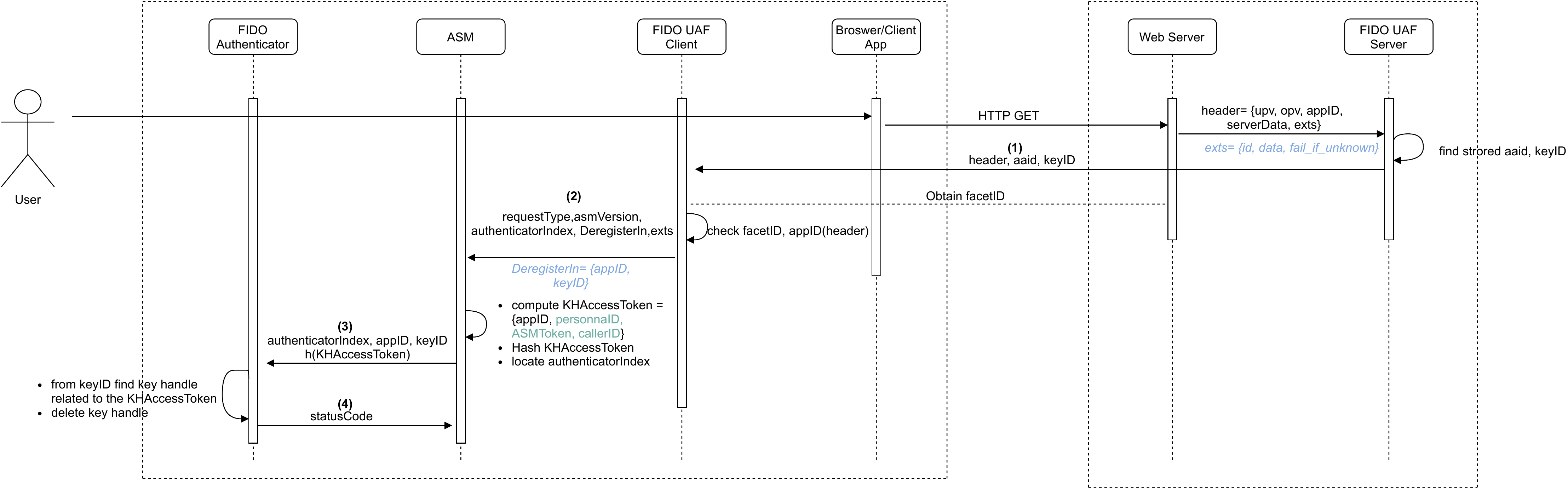}
\caption{FIDO UAF Deregistration Message Flow}
\label{fig:UAF_Deregistration_flow}
\end{figure}

\begin{enumerate}
\item \textbf{Deregistration Request:}
The deregistration process is required when the user account is removed from the relying party. The later can trigger the deregistration by asking the authenticator to delete the associated UAF credentials that are bound to the user account. The flow is illustrated on Fig.~\ref{fig:UAF_Deregistration_flow}. To achieve that, the FIDO UAF Server will send a message to the FIDO UAF Client containing the 
\begin{itemize}
    \item the \textit{header} (upv, op, appID, serverData, exts)
    \item the information on the authenticator to be deleted, signifying the \textit{aaid} and the \textit{keyID}
\end{itemize}
\item \textbf{Deregister ASM Request:}
The FIDO Client will create an ASM Request 
\begin{itemize}
    \item the \textit{requestType} which is set to “deregistration”
    \item \textit{asmVersion}
    \item \textit{authenticatorIndex}
    \item \textit{args} of “DeregisterIn” which contains the \textit{appID} and the \textit{keyID}.
\end{itemize}
\item \textbf{Deregister Command:}
The ASM will locate the authenticator using the \textit{authenticatorIndex}, construct the \textit{KHAccessToken} and  afterwards it will send to the authenticator a message which includes
\begin{itemize}
    \item the \textit{authenticatorIndex}
    \item the \textit{appID} (optionally)
    \item the \textit{keyID}
    \item the hashed \textit{KHAccessToken} (appID and if the authenticator is bound personnaID, ASMToken, callerID)
\end{itemize}
\item \textbf{Deregister Response:}
The authenticator will delete the keys related to the specific \textit{appID} and user account and it will send back to the ASM 
\begin{itemize}
    \item a \textit{status code} indicating success or failure
\end{itemize}
\end{enumerate}

\subsection{Differences between versions 1.0, 1.1 and 1.2 }
\label{subsec:UAF_differencesBetweenVersions}
The first generation of FIDO UAF specifications, 1.0 defined Basic Full and Basic Surrogate as attestation types. ECDAA algorithm was introduced for the first time in v1.1. Modifications have been introduced in v1.2 though, in order to overcome the Diffie-Hellman oracle of TPM \cite{7958616}.
\par Since the upv and the asmVersion parameters indicate the supported UAF version, their value is altered in different versions of the protocol. Additionally, more codes were added in the authenticatorType, the attestationType, the authenticationAlgorithm and the statusCode dictionaries between v1.0 and v1.1. The length of the display characteristics information was attached in the display characteristics parameter. In v1.2 the Metadata Statement includes more information than the previous version such as the alternativeDescriptions, operatingEnvironment as well as the cryptoStrength. Additions were also applied in the authenticatorStatus in order to include more information regarding the certification. The extensions parameter has been also more thoroughly explained in the latest specifications.

\subsection{Advanced Security}
\label{subsec:Advanced_sec_UAF}
\par
Alike FIDO2, UAF provides protection against phishing attacks by linking keys to relying parties (using the appID and channelBinding information). The randomly generated challenge provides protection against replay attacks while UAF provides round-trip integrity. The communication between client and server is performed via https protocol. Additionally, UAF offers a wide choice of options regarding the security policy criteria. 
Apart from User Authentication, Transaction Confirmation is also covered, leveraging of What You See Is What You Sign (WYSIWYS) option offered by a Trusted User Interface (TUI), fact that makes the protocol a perfect candidate for online marketplaces.
The signature counter provides synchronisation as well as resilience against cloned authenticators. 

\subsection{FIDO UAF vs FIDO2}
\label{subsec:FIDO UAF vs FIDO2}
Both FIDO UAF and FIDO2 offer passwordless authentication. Nevertheless there are some differences between the two. The obvious differences is that each protocol serves a different scope. While FIDO2 aspires to be a web framework dedicated to web browsers via W3C, FIDO UAF is mainly dedicated to the mobile experience. On the other hand there are differences introduced from past experience. For example, FIDO2 defines a \textit{timer} for the client timeout of the request while the notion of token binding as well as ECDAA attestation is not included probably due to implementation difficulties and limited acceptance from the developer's side. In the place of ECDAA, Privacy CA has been introduced as a solutions that offers privacy to the user but is also easier to implement. The latest CTAP2 specifications mention fingerprint enrollment. In general, the issue of privacy seems to be of major importance for WebAuthn since the specifications dedicate a section to privacy enhancing mechanisms. Compatibility between UAF and FIDO2 is achieved though through the \textit{clientData} object which is also included in the latest UAF v1.2 specifications as a substitute to \textit{fcParams}. 
\par
On the other hand, UAF seems to be more specific when it comes to the communication of the client application with the authenticator in order to identify all available authenticators \textit{GetInfo} while the notion of bound and silent authenticators is explicitly defined.  Extensions in UAF's versions v1.0 and v1.1 are processed by the client while in UAF v1.2 and FIDO2 the extensions are also processed by the authenticator through the \textit{clientData} object. Additionally, UAF specifies two different \textit{counters}: one for i) registration (attestation) and another for ii) authentication (assertion) while the server's \textit{policy} includes various criteria compared to FIDO2 plain \textit{AAGUID} which in UAF is equivalent to the \textit{AAID}. UAF defines the notion of personna which in FIDO2 is not discussed. Lastly UAF does not offer to the client the option to choose a different hash algorithm between registration and authentication. A summary of the major differences between FIDO UAF and FIDO2 is demonstrated on Table~\ref{tab:Differences FIDO UAF vs FIDO2}.
\begin{table}[t]
\centering
\caption{Differences between FIDO UAF and FIDO2.}
\label{tab:Differences FIDO UAF vs FIDO2}
\scriptsize
\begin{tabularx}{\linewidth}{L|L}
\toprule
\textbf{FIDO UAF} &\textbf{FIDO2}\\
\midrule
Usually developed as an SDK &API\\\hline
Mostly used for mobile applications &Used for the Web \\\hline
appID as well as facetID for every “subdomain” or “facet” of the app (i.e., Android, iOS ). appID is RPID in FIDO2 &Origin to facilitate the implementation as in UAF every subdomain (due to implementation difficulties) would need separate registration \\\hline
Limited support &Wider support, more open source works, browsers and OS have implemented their side\\\hline
The timeout refers to the authenticator, if the authenticator takes too long then it will return a timeout error message to the client. Nevertheless the client is not instructed to set a desired timeout time. A timer does not exist for the client. &The client initiates a Timer related to the timeout of the request \\\hline
Specify federation namely  &Does not specify federation namely and describes that future work is needed in order to specify the tokens to add some form of manifest format with properties that declare the authentication type which the provider supports (in credential management document not in the WebAuthn). \\\hline
Fingerprint enrolment was not part of the UAF spec. &Fingerprint enrolment among with other parameters such as token PIN is described in CTAP2 \\\hline
Based on the notion of token binding  &No token binding up to this day\\\hline
ECDAA was the major option &ECDAA was rejected in the WebAuthn L2 specs since it was never implemented \\\hline
The specifications define Basic surrogate &Privacy CA was added\\\hline
Did not take the issue of privacy under consideration in the first versions of UAF specs &WebAuthn has a separate section for privacy consideration (i.e., deanonymisation techniques)\\\hline
In the first versions of UAF fcParams (appID; challenge; facetID; channelBinding) was only the parameter for validation while clientData (challenge; origin; hashAlg; tokenBinding; extensions) was added in 1.2 \begin{itemize}
    \item 	This challenge in clientData plays a similar role as the challenge field in FinalChallengeParams.
    \item This origin plays a similar role as the the facetID field in FinalChallengeParams.
    \item The hashAlg allows the client can freely select the hash algorithm - unlike FinalChallengeParams, where the authenticator must use the same algorithm as for signing the assertion. 
    \item This tokenBinding in clientData plays a similar role as the channelCinding field in FinalChallengeParams. 
\end{itemize}
 &clientData is the only the parameter for FIDO2\\\hline
FIDO UAF defines every communication (i.e., \textit{Get Info, Transaction Confirmation} and \textit{Deregistration}) &FIDO 2 does not define communication between browser and OS or the OS with the authenticator \\\hline
UAF clearly explains what is saved in the authenticator and in the ASM  &WebAuthn mostly analyses what is stored in the authenticator\\\hline
Silent authenticator are discussed &Silent authenticators are not discussed but do exist when it comes to disable them\\\hline
Extensions in older versions (1.0 and 1.1) were processed only by the client. &In FIDO 1.2 and FIDO2 extensions, through the \textit{clientData} stracture the extensions can be also processed by the authenticator\\\hline
Registration and sign counter &Sign counter only. Possibly because the registration counter was not so meaningful in terms of security.\\\hline
The server’s policy includes acceptance criteria with many subcategories &More “relaxed”  using the AAGUID to exclude authenticators\\\hline
AAID &AAGUID\\\hline
Defines the notion of personna &Clearly defined how the user can choose among different accounts at the same site\\\hline
In v1.0 and v1.1 the fcParams the client cannot choose the hash algorithm as the authenticator must use the same algorithm used during registration. In v1.2 due to the addition of the clientData, the client may also choose the hash algorithm. &The client may select the hash algorithm \\
\bottomrule
\end{tabularx}
\end{table}

\section{FIDO U2F/CTAP1}
\label{sec:U2FandCTAP1}
One Time Passwords (OTP) are not considered safe alternatives compared to U2F, since numerous attacks have been launched over the last years. SMS OTP have proven a significant and simple attack vector due to the cellular network’s vulnerabilities. FIDO’s Universal Second Factor Authentication (U2F) protocol allows online services to augment the security of their existing username and password scheme by adding a strong second factor. The user after providing the username and password to the requested service, will also have to present an additional information (i.e., a device). U2F’s focus on defining the JavaScript API which allows the communication between the server, the FIDO client (i.e., browser) and the authenticator. The FIDO U2F supports two operations: i) registration and ii) authentication.

\subsection{U2F Registration}
\label{subsec:U2FRegistration}
\begin{figure}[t]
    \centering\includegraphics[width=0.7\textwidth]{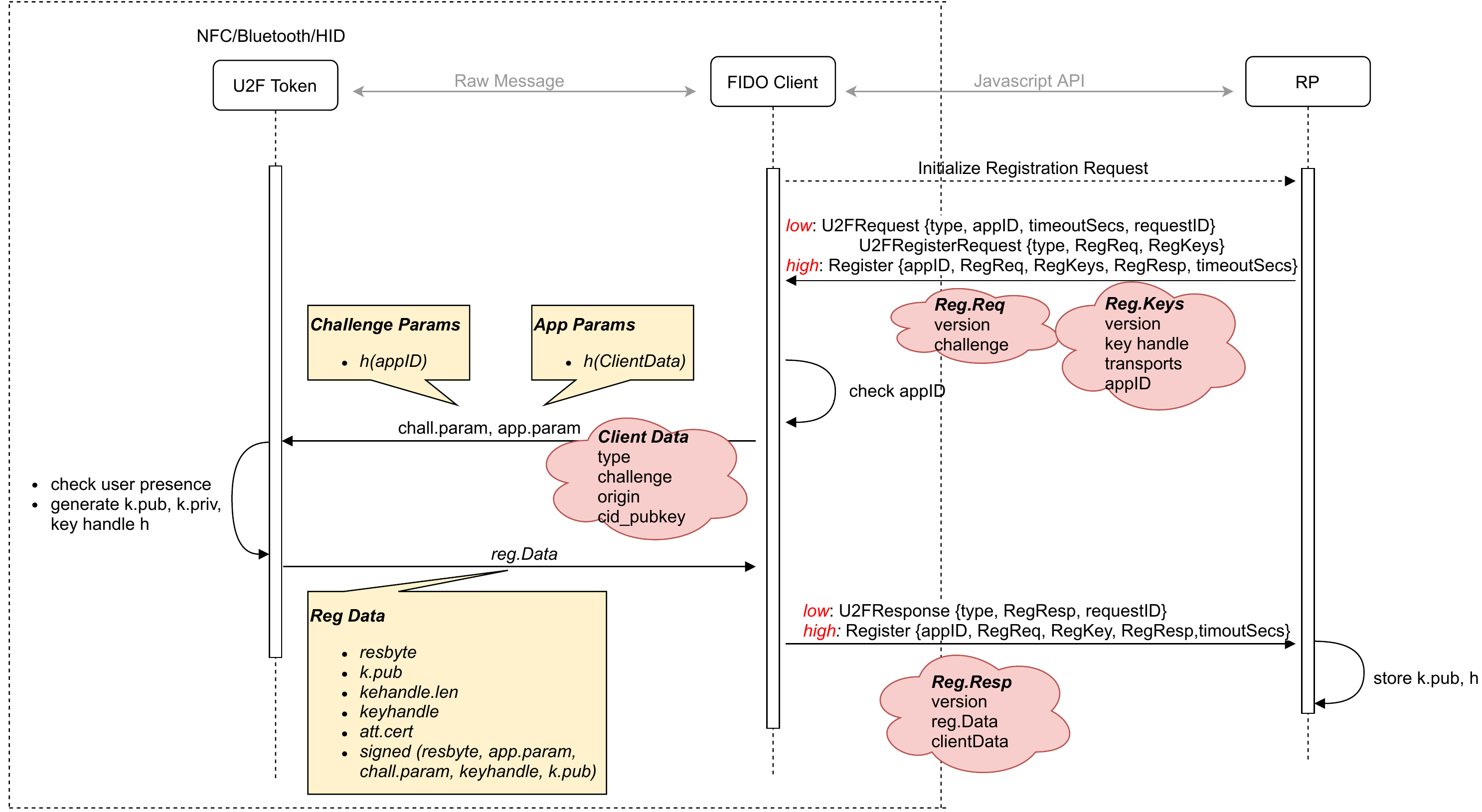}
    \caption{U2F Registration}
    \label{fig:u2f_reg}
\end{figure}

The communication between the FIDO client and the relying party starts with the first requesting to register. The U2F API may be exposed to web pages either on low or high level, which is built on top of the MessagePort API. It is generally recommended to use the high-level JavaScript API. Depending on the implementation of the API, some of the exchanged messages differ. Nevertheless, the relying party performs some common operations, as depicted in Fig.~\ref{fig:u2f_reg}.
\begin{enumerate}
    \item \textbf{U2f Register Request: }The relying party creates the Registration Request which implies that it decides the U2F version to be used and the appID. Additionally, the relying party generates the random challenge and stores all the information related to the registration request. 
    \item \textbf{MessagePort API U2F Request}
    \begin{itemize}
\item \textbf{Option\_1: Low-level} The relying party sends to the FIDO client the U2fRequest containing the i) type of the request implying a \texttt{u2f\char`_register\char`_request} in this case, ii) the appID, iii) the timeoutSeconds and iv) the requestID. The \texttt{u2f\char`_register\char`_request} dictionary further contains the v) register requests which indicate  the version and the challenge and the vi) registered keys which indicate the version, the key handle, the transports (Bluetooth, NFC, USB) and the appID related to the specific key handle. 

\item \textbf{Option\_2: High-level}The FIDO client upon receiving the registration request (either on low-level or on high-level MessagePort API), creates the client data object which includes the i) type, ii) the server challenge, iii) the origin and iv) the public key of the TLS channel. Afterwards it generates the challenge parameter which is the hashed client data object using SHA-256 and sends it among with the application parameter, which is the hashed appID, to the U2F device.\end{itemize}
\item \textbf{U2F Register Raw Message: }The U2F device after receiving the message from the FIDO client, ensures the user’s presence and generates a key pair (k.pub and k.priv) and the corresponding key handle which facilitates the detection of the key pair if there are several. Next, the U2F device creates the response message which is addressed to the FIDO client and contains i) a reserved byte which is set to 0x05 for legacy reasons, ii) the public key, iii) the length of the key handle, iv) the key handle, v) the attestation certificate and vi) the signature of the reserved byte, the application parameter, the challenge parameter, the key handle and the user’s public key.
\item \textbf{U2f Register Response: }
\begin{itemize}
\item \textbf{Option\_1: Low-level MessagePort API U2F Response } The FIDO client creates the U2f response U2fResponse including i) the type of message, which is register, ii) the response data, containing the version of the U2F protocol supported by the device, the registration data that is the U2F register raw message (as described in 4.1.2) and the client data object and iii) the requestID . The U2f response is send from the FIDO client to the relying party.
\item \textbf{Option\_2: High-level MessagePort API U2F Response }
This API provides a u2f object with an interface for the register and sign operations. The interface for register contains i) the appID, ii) the register request (version, challenge), iii) the registered key (version, key handle, transports, appID), iv) the register response containing the version of the U2f device, the U2f register raw message and the client data or an error handler and v) the timeout seconds.The relying party after receiving the registration response either accepts or declines the registration. \end{itemize}
\end{enumerate}

\subsection{Authentication}
\label{subsec:U2F_Authentication}
\begin{figure}[t]
    \centering
    \includegraphics[width=0.7\textwidth]{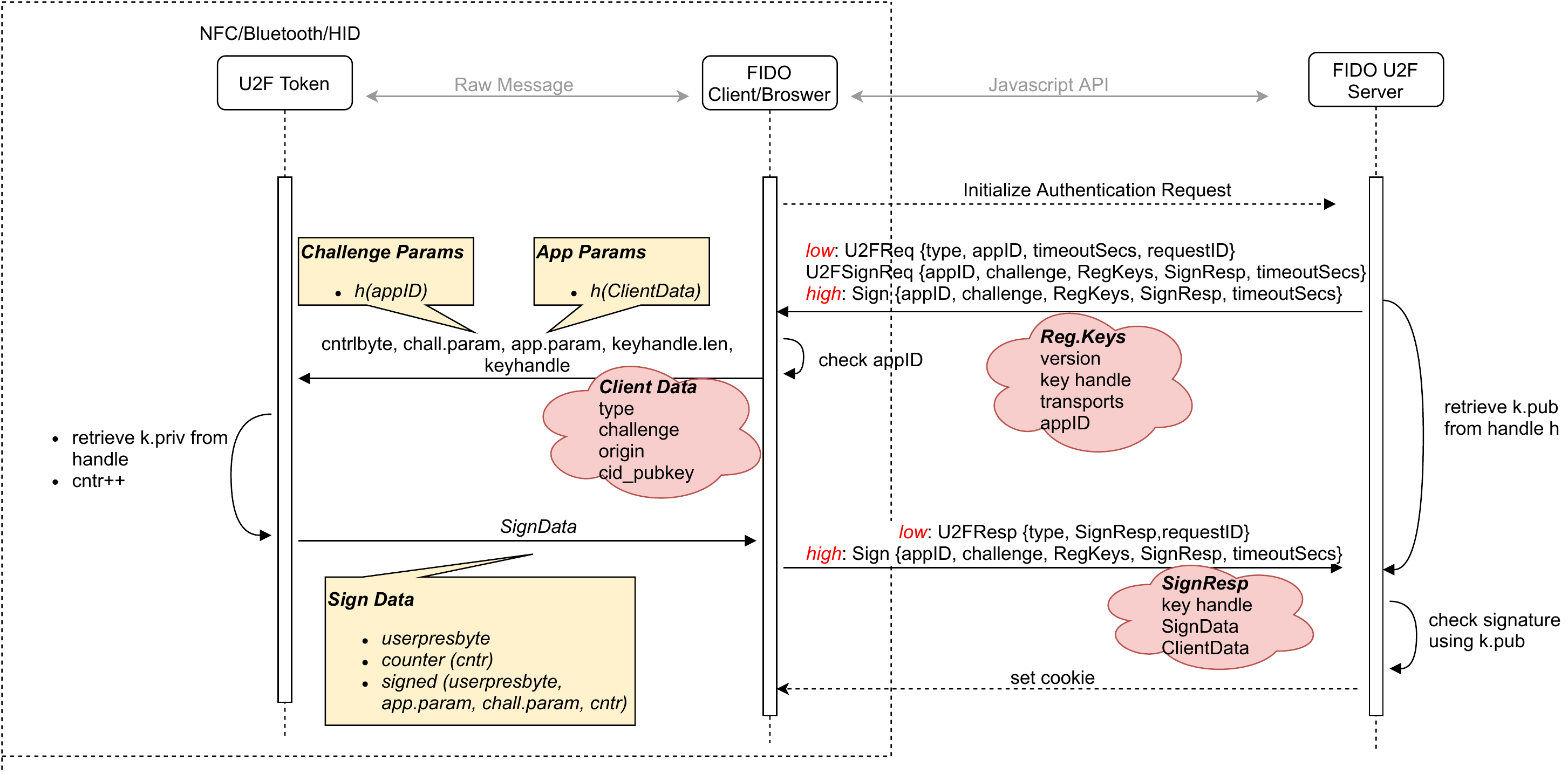}
    \caption{U2F Authentication}
    \label{fig:u2f_auth}
\end{figure}

Similarly to the registration, the authentication features the low and the high-level definition of the API. The detailed flow is showcased in Fig.~\ref{fig:u2f_auth}.
\begin{enumerate}
    \item \textbf{MessagePort API U2f Sign Request} The relying party sends a sign request to the FIDO client in order to authenticate the user. As presented for the registration process, the FIDO client may choose to support either the low-level or the high-level MessagePort API. 
    \begin{itemize}
        \item \textbf{Option\_1: Low-level } The relying party sends to the FIDO client the U2fRequest containing the i) type of the request implying a \texttt{u2f\char`_sign\char`_request} in this case, ii) the appID, iii) the timeoutSeconds and iv) the requestID. The \texttt{u2f\char`_sign\char`_ request} dictionary further contains v) the server’s challenge and the vi) registered keys which indicate the version, the key handle, the transports (Bluetooth, NFC, USB) and the appID related to the specific key handle.
        \item \textbf{Option\_2: High-level }This API provides a u2f object with an interface for the register and sign operations. The interface for register contains i) the appID, ii) the challenge, iii) the registered keys (version, key handle, transports, appID), iv) the sign response which is null at this point and v) the timeout seconds.
        \par
        The FIDO client verifies the appID of the caller and uses the challenge to create the client data object. Later, it constructs the authentication request message destined to the U2F device. This message contains i) a control byte, ii) the challenge parameter, which is the hash of the client data, iii) the application parameter that is the hashed appID, iv) the key handle and its length. 
    \end{itemize}
    \item \textbf{U2f Authenticate Raw Message: } The U2F device upon receiving the authentication request creates the authentication response message which includes i) a user presence message, ii) a counter and iii) the signature of the application parameter, the user presence byte, the counter and the challenge parameter.  In parallel, the U2f device increments the counter in each authentication operation. 
    \item \textbf{U2f Sign Response: }The FIDO client after receiving the message from the U2F device, creates the sign response addressed for the relying party which contains i) the key handle, ii) the signature data that refers to the U2F device’s raw message (as described in 4.2.2) and iii) the client data. 
\end{enumerate}

\subsection{Advanced Security}
Similarly to FIDO2 and FIDO UAF, U2F checks the validity of the authenticator as well as, the relying party information therefore round-trip integrity is achieved in U2F as well. The randomly generated challenge protects from replay attacks while the counter protects from cloned authenticators. The main difference that only user presence is needed while the relying party does not define the policy as in UAF. 

\section{Market penetration and applicability}
\label{sec:MarketSurvey}
\subsection{Survey methodology}
\label{subsec:survey_meth}
The work in \cite{Support_for_FIDO2WebAuthnCTAP} and \cite{CAN_I_USE_WEBAUTHN} offer an overview the browsers that support WebAuthn. It is evident that all major browsers (i.e., Mozilla, Chrome, Opera, Safari and Edge) provide WebAuthn support at least for their desktop version. 
Regarding the underlying operating systems, Microsoft's desktop version is currently the one with the wider support allowing all types of authenticators. MacOS 14 enables users to use Touch ID and Face ID for web logins, while it supports external authenticators as well with the exception of NFC protocol that is not supported by macOS browsers in general, and Firefox CTAP2 support which is not yet compatible. For in Linux FIDO is available only through USB authenticators \cite{yubico_webauthn_browsers}, \cite{apple_ios14}.
Mobile platforms such as Android 7.0 (“Nougat”) is certified as compatible with FIDO2 since February 2019 \cite{FIDO2_Android_OS}, which means that most of the recent Android devices are FIDO-ready supporting CTAP2 for both bound/internal and external  authenticators, with the only exception of NFC in Firefox. In iOS, WebAuthn and CTAP2 are offered only Safari browser with the exception of BLE for authenticator transport protocol.
\par
In \cite{USB-Dongle_Authentication} and \cite{Works_With_Yubikey} services and websites that leverage of WebAuthn and U2F are enlisted. Although the majority of the aforementioned services have yet to completely opt-out of the username and password scheme, FIDO constitutes a popular second factor authentication option. To this day though, there are not many options from completely passwordless authentication. The most widely-adopted passwordless experience has been rolled out by Microsoft, supporting both \textit{User Presence} and \textit{User Verification} as well as \textit{Resudenty Key}\cite{yubico_webauthn_browsers}. The Windows 10 Hello for Business allows corporate users to authenticate to an Active Directory or Azure Active Directory using FIDO2 passwordless authentication. The user has priory enrolled to Windows Hello through a two-step verification method. After the initial enrollment the user may authenticate to the device using a passwordless method such as a gesture or a fingerprint. Microsoft advises administrators, apart from identifying the password usage and plan and mitigation, to define the different work personas as a mean to manage the different levels of privileges between the multiple departments of an organisation. As described in \cite{microsoft_azure_mfa}, Windows Hello lets users authenticate to:
\begin{itemize}
    \item a Microsoft account
    \item an Active Directory account
    \item a Microsoft Azure Active Directory (Azure AD) account
    \item Identity Provider Services or Relying Party Services that support FIDO v2.0 authentication (in progress)
\end{itemize}

Samsung Pass, using Samsung Knox, is a FIDO-enabled service that enables strong authentication across applications using biometrics and combined with a cloud-based service provided by Samsung \cite{samsung_knox_android}. With Samsung Pass, smartphone users can lock up multiple sets of authentication credentials (from both public and private enterprise services) with biometrics \cite{samsung_pass}. It is not clearly elaborated though whether Samsung Pass supports FIDO2 or UAF or both.
\par
The existing literature does not cover the UAF's level of penetration from the industry. In order get an indication of UAF's applicability, we chose to investigate the FIDO UAF certified showcases with a commercial deployment, as appeared in \cite{fido_showcase}. In \cite{li2020authenticator}, researchers have chosen a set of applications to examine for UAF nevertheless, their dataset was focused on the Asian market, while it did not include the certified showcases, use cases or products that are enlisted as FIDO2/U2F enabled. The main objective was to prove whether FIDO UAF certified clients are still supported or they have been replaced by FIDO2/WebAuthn, and answer the question of whether FIDO UAF has become obsolete, or not.
\par
The survey focused on commercial, FIDO UAF certified, applications available at Google Play Store, as Android is the most popular operating system in mobile devices at a global scale \cite{MobileOS_Market_Share_Worldwide}, while there are more open-source tools available for reverse engineering the applications (i.e., \textit{apktool}). Additionally, regarding native application development, due to the openness of the Android platform in comparison with the iOS, it was presumed as more a likely scenario to find Android applications that support FIDO as i) there are more projects available as examples of a FIDO UAF Clients on GitHub and dated older than the respective iOS ones \cite{github_ebay_uafclient} and ii) Android was the first platform to offer FIDO2 OS-level support \cite{FIDO2_Android_OS}. Similarly to previous work \cite{10.1145/3291533.3291573}, our methodology for reverse engineering and processing the applications was the following: 
\begin{enumerate}
    \item Acquire the \textit{.apk} file of the application to-be-investigated
    \item Use a\textit{pktool} and process the AndroidManifest.xml to discover any dependencies with FIDO
    \item Use \textit{dex2jar} tool to investigate for FIDO in the .jar classes 
\end{enumerate}
The survey which spanned a period from November to May 2021 was aided by the \textit{reverse\_apks} \footnote{\url{https://github.com/AnnaAnge/reverse\_apks}}, which automates the above-mentioned procedures. Although the research methodology, applied for the survey, returns results regarding the applications found to support FIDO with high certainty, it is different for applications in which FIDO was not discovered. This is due to two main reasons, namely,i) the \textit{.apk} file may have mechanisms that prevent reverse engineering and ii) the application may support FIDO but through third-party libraries. In particular, the \textit{.apk} files were acquired from either \textit{apkmonk} \footnote{\url{https://www.apkmonk.com}} or \textit{apkpure} \footnote{\url{https://apkpure.com}} websites. 
\par
In addition to the enlisted services in \cite{USB-Dongle_Authentication} and \cite{Works_With_Yubikey}, the certified products \cite{fido_showcase} and the performed reverse engineering of Android applications, we ran a keyword search for the FIDO protocols, utilising the GitHub API, in order to identify the level of acceptance as well as the languages preferred by developers. The statistical results from the keyword search queries that we applied are illustrated in Table Tab.~\ref{tab:github_repos}. From the retrieved results, it is safe to assume that there is active interest on the FIDO protocols from the developer's side. Currently there are 331 and 135 public repositories that include the word "FIDO" and "FIDO2" respectively in their description. Delving into more detail, WebAuthn seems to be the leading FIDO protocol (295 public repositories) with U2F behind (86 public repositories) followed by CTAP (26 public repositories) and UAF (17 public repositories).

\subsection{FIDO applicability}
\label{subsec:FIDO_applicability}
FIDO2/WebAuthn and U2F covers a wider market. Among digital services that use FIDO, Google (U2F and FIDO2), Dropbox, Github, Twitter, Yahoo Japan are included. Usually FIDO2 is used as a second factor authentication, as an upgrade to a pre-existing U2F implementation. Passwordless authentication via FIDO2 or UAF was found to be less common, with the exception of Windows Hello login.
\par
UAF was discovered to be a less popular option. From our analysis we concluded that i) Paypal, ii) Line Pay, iii) SoftBank, iv) Bank of America, v) Bank of China, vi) Shinhan Bank, vii) Revolut, viii) National Health Service (NHS) ix) T-Mobile, x) NokNok, xi) HYPR, xii) ReCred \footnote{\url{https://www.recred.eu/}}, xiii) 1Password, xiv) Google Play, xv) Cloudflare, xvi) 11st and xvii) TW FidO supported FIDO in their mobile applications. More specifically, Bank of America and Line Pay supported FIDO2 while the rest supported FIDO UAF protocol. Therefore, it was presumed that for Financial Services \& Banking as well as Online markets \& Retail that require a higher level of security both for authentication and for transaction confirmation, UAF was the preferred option, in the mobile application version. Interestingly enough, NHS which belongs in the Health sector was also found to support UAF, indicating that applications that handle sensitive data, trust UAF. 


\subsection{Survey of FIDO-related scientific publications}
\label{subsec:literature_review}
In order to identify and analyse the research works with interest to FIDO, a survey was performed leveraging of the dblp API~\footnote{\url{http://dblp.org/search/publ/api}}, for six keywords: i) FIDO, ii) FIDO2, iii)UAF, iv) U2F, v) WebAuthn and vi) CTAP. The retrieved results from the dblp API are illustrated in Fig.~\ref{tab: DBLP_results_per keyword}. We have further processed the results in order to exclude duplicates. For example if a paper contained both "FIDO2" and "WebAuthn" in its title, we kept it only once (i.e., in WebAuthn since it was more specific). After the initial process of the results, we manually identified whether the title "FIDO"  or "FIDO2" was dedicated to specific version of the protocol.
\par
Judging from the retrieved results, "FIDO" and "FIDO2" is the most commonly found keyword in titles. Nevertheless after thoughtful study on the documents, U2F was found to be the most popular FIDO protocol among research works followed by UAF and FIDO2. Additionally, most research works have been published for the first version of the FIDO protocols than the second. 
\begin{table}[t]
\centering
\caption{Numerical results from the keyword search in DBLP database}
\label{tab: DBLP_results_per keyword}
\begin{tabular}{cccccc}
\toprule
\textbf{CTAP} &\textbf{WebAuthn} &\textbf{U2F} &\textbf{UAF} &\textbf{FIDO2} &\textbf{FIDO} \\
\midrule
0 &3 &7 &6 &6 &1\\
\bottomrule
\end{tabular}
\end{table}

This result is particularly interesting  considering the deviation between the research/academic field and the industry. While the first version of the protocols (i.e., U2F and UAF) is broadly investigated by academia, in the commercial world, the second version of the protocols (i.e., WebAuthn and CTAP) is the most prominent, as in Section~\ref{subsec:implementation_difficulties}. It has to be underlined that CTAP was not found as the main subject in any research work.
\begin{enumerate}
    \item FIDO - In \cite{9346319} FIDO is used for user authentication in a smart energy environment.The exact version of FIDO is not implicitly specified.   
    \item UAF - Both \cite{fengformal} and \cite{10.1007/978-3-319-67639-5_11} provide an analysis of the UAF protocol from a cryptographic and security perspective respectively. In \cite{li2020authenticator} an attack against the protocol is performed whereas \cite{9045440} and \cite{8947083} explore the protocol's applicability, leveraging of UAF in order to provide authentication as well as identification.
    \cite{a37db6c69c174594bb7fe0c87ceca824} discusses a TEE-based implementation of the UAF's Transaction Confirmation feature. 
    \item U2F - The authors in \cite{cryptoeprint:2020:1298} performed a study on user's susceptibility on phishing attacks even with FIDO enabled. The notion was that FIDO could be downgraded to weaker 2FA alternatives. In \cite{Of_two_minds_with_2factor_238317},  \cite{10.1007/978-3-662-58387-6_9} and \cite{238325} usability studies of U2F with external participants are performed. A U2F based protocol is presented in \cite{10.1145/3063955.3063982} for mobile payments using the mobile phone's UMTS Subscriber Identity Module (USIM) and Secure Element (SE) while in \cite{7860546} an extension of U2F is proposed in order to support smart cards and mobile phones as authenticators. \cite{cryptoeprint:2017:721}
    \item FIDO2 - The work in \cite{barbosaprovable} provides a cryptographic security analysis of the protocol, while \cite{9343231} proposes an extension to the protocol in order to enable continuous authentication. In \cite{FIDO2Kingslayer} and \cite{255646} usability studies are presented. \cite{wagner2021remote} and \cite{9343176} suggest remote and external FIDO2 authentication respectively, using QR codes. In \cite{10.1145/3319535.3363258} a sim TPM FIDO authenticator is presented. 
    \item WebAuthn - A theoretical framework for a remote FIDO2 authentication, in cases where specialised hardware is unavailable is presented in \cite{wagner2021remote}. Authors in \cite{10.1145/3319535.3363283} focus on the pitfalls of WebAuthn for developers indicating that lack of sufficient understanding of the protocol,  insecure or incomplete of libraries and privacy concerns are pivotal issues for the community. A novel cryptographic primitive is proposed in \cite{frymann2020asynchronous} in order to enable asynchronous key generation as well as account recovery for WebAuthn.  
    \item CTAP - To this day, there is no research work solely devoted on CTAP. 
\end{enumerate}

\begin{table}[ht]
\centering
\caption{Summary of business sectors and companies that use FIDO}
\label{tab:vulnerabilities}
\scriptsize
\begin{tabularx}{\linewidth}{lllll}
\toprule
\textbf{Business Sector} &\textbf{Company} &\textbf{Country} &\textbf{Web App} &\textbf{Mobile App}\\
\midrule
\multirow{3}{*}{Backup \& sync} &Dropbox &US &\checkmark &X \\
&Files.com &US &\checkmark &N/A \\
&Google Drive &US &\checkmark &X \\
\hline
Cloud Computing &Amazon Web Services, Google Cloud Platform &US &\checkmark &X \\
\hline
\multirow{3}{*}{Cryptocurrencies} &Bitfinex &Hong Kong &\checkmark &X \\
&Coinbase &US &\checkmark &X \\
&Gemini &US &\checkmark &X \\
\hline
\multirow{5}{*}{Developer} &Bitbucket &Australia &\checkmark &N/A \\
&Github &US &\checkmark &X \\
&Gitlab &US &\checkmark &N/A \\
&PyPI & &\checkmark &N/A \\
&Sentry &US &\checkmark &N/A \\
\hline
Domains &Gandi, Namecheap, Porkbun  &US &\checkmark &X \\
\hline
\multirow{2}{*}{Education} &University of Miami &US &\checkmark &X \\
&SUNET (University Computer Network) &Sweden &\checkmark &N/A\\
\hline
\multirow{2}{*}{Email} &Fastmail, Gmail, Yahoo &US &\checkmark &X \\
&Mail.de, Tutanote &Germany &\checkmark &X \\
\hline
Entertainment &Google Play &US &\checkmark &\checkmark\\
\hline
\multirow{10}{*}{Financial Services \& Banking} &Bank of America &US &\checkmark &\checkmark FIDO2\\
&Bank of China &China &X &\checkmark UAF 1.0\\
&Google Pay &US &\checkmark &X\\
&Line Pay &Japan &\checkmark &\checkmark FIDO2\&U2F\\
&Paypal &US &TBD &\checkmark\\ 
&Revolut &US &X &\checkmark does not open .jar\\
&Shinhan Bank &Korea &X &\checkmark \\
&SoftBank Corp. &Japan &X  &\checkmark\\
&Stripe &US &\checkmark &X \tiny{(not in .xml, jar was obfuscated)} \\
\hline
\multirow{7}{*}{Government} &gov.uk &UK &\checkmark &X\\
&login.gov &US &\checkmark &N/A\\
&Ministry of Interior (TW FidO) &Taiwan &No info &\checkmark\\
&Electronic Transactions Development Agency (ETDA) &Thailand &No info &TBD \cite{fido_government_deployments}\\
&Canadian Digital Service (CDS) &Canada &\checkmark &N/A \\
&Korea National Intelligence Service (KCIA) &Korea &\checkmark &N/A \\
&CZ.NIC &Czech Republic &\checkmark &N/A \\
&SUNET (eduID) &Sweden &\checkmark &N/A \\
\hline
\multirow{2}{*}{Health and insurance} &National Health Service (NHS) &UK &TBD \cite{fido_nhs} &\checkmark\\
&Google Fit &US &\checkmark &X \\
\hline
\multirow{2}{*}{Hosting} &Nitrado &Germany &\checkmark &X\\
&OVH &US &\checkmark &X \\
&GoDaddy  &US &\checkmark &X \\
&Opalstack &US &\checkmark &X \\
\hline
\multirow{5}{*}{Identity Management} &1Password &US &\checkmark &\checkmark FIDO2\&U2F \\
&Bitium &US &\checkmark &N/A \\
&Bitwarden &US &\checkmark &X \\
&Dashlane &US &\checkmark &X \tiny{(not in .xml, jar was obfuscated)} \\
&Keeper &US &\checkmark &X \\
&RSA SecurID Access &US &\checkmark &X \\
&okta &US &\checkmark &X \\
\hline
Investing &Vanguard &US &\checkmark &N/A\\
\hline
\multirow{3}{*}{Online markets / Retail}
&11st &US & &\checkmark UAF 1.0\\
&Shopify &Canada &\checkmark &X \tiny{(not in .xml, jar was obfuscated)} \\
&Ebay &US &\checkmark &Neither .xml or .jar files were readable\\
\hline
Other &Office365 &US &\checkmark &X \\ 
\hline
\multirow{5}{*}{Security} &Boxcryptor &Germany &\checkmark &X \\
&Cloudflare &US &\checkmark  &\checkmark UAF 1.0\\
&dmarcian &US &\checkmark &N/A\\
&HYPR &US &\checkmark &\checkmark \\
&NokNok &US &\checkmark &\checkmark \\
\hline
\multirow{8}{*}{Self-hosted} &Authelia & &\checkmark &N/A \\
&Gluu &US &\checkmark &N/A \\
&GreenRADIUS &US &\checkmark &N/A \\
&LinOTP &Germany &\checkmark &X \\
&Nextcloud &Germany &\checkmark &X \\
&phpMyAdmin & &\checkmark &N/A \\
&privacyIDEA & &\checkmark &N/A \\
&Wordpress (software) &US &\checkmark &X \\
\hline
\multirow{2}{*}{Social media} &Facebook &US &\checkmark &X\\
&Twitter &US &\checkmark &X\\
\hline
Research Projects &H2020-DS-02-2014-ReCred &EU &X &\checkmark UAF 1.0\\
\hline
\multirow{2}{*}{Telecom \& Mobile} &NTT DOCOMO &Japan &X &X (samsung auth sdk)\\
&TMobile &US &X &\checkmark\\
\hline
\multicolumn{4}{l}{\textit{\checkmark:FIDO Enabled, X:FIDO not found, N/A:Web version only, TBD:To Be Developed}}\\
\bottomrule
\end{tabularx}
\end{table}

\section{Critique on FIDO protocols adoption}
\label{sec:critique}
The critique on FIDO adoption across the various markets, according to the performed survey, is organised around 4 main pillars. The first pillar of the analysis includes the developer perspective. Specifically, the discussion is focusing on the aspects that render UAF and FIDO2 more popular and developer friendly (i.e., wider support from the open source communities, etc.,) for the service developers and providers. Second pillar of the analysis is concentrated on the security shortcomings and known vulnerabilities that still remain open. The third pillar is focusing on the emerging markets and services in the Next Generation Internet and 6G era. Such landscape includes areas such as IoT+AI, Industry 4.0 and beyond, etc. More complex privacy requirements (AI-based decision making for sensitive services: healthcare, personal data, etc.).

\subsection{Compliance to standards and regulations (NIST 800-63, KYC, PSD2, PCI-DSS and GDPR}
\label{subsec:NIST_psd2_GDPR}
\begin{figure}[ht]
\centering
\includegraphics[width=0.5\textwidth]{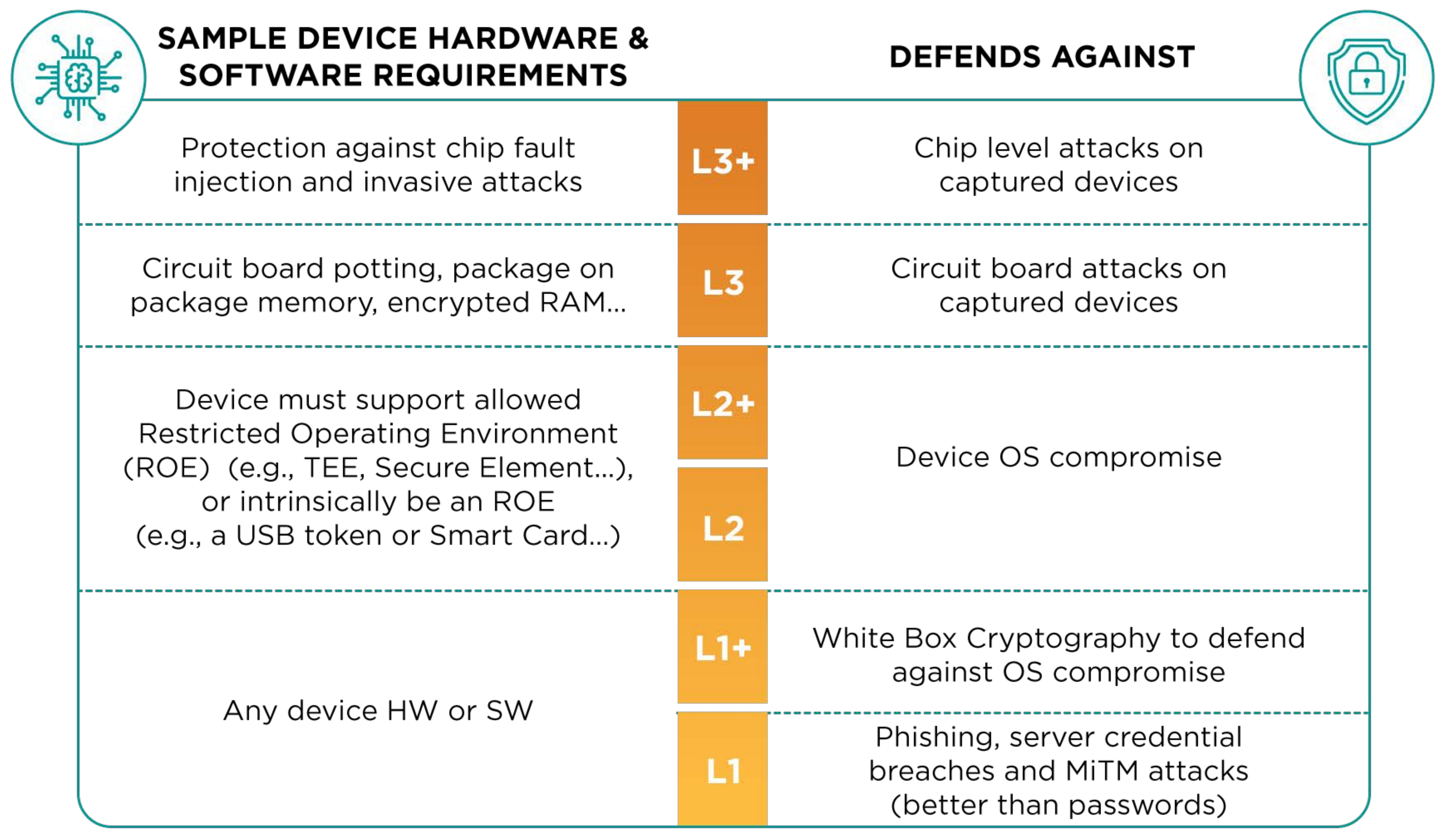}
    \caption{FIDO requirements according to certification levels per NIST}
    \label{fig:FIDO_cert_and_levels}
\end{figure}

Arguably, cybersecurity attacks may have irreversible consequences for the vitality of an organisation therefore usually  security thus privacy requirements are dictated by regulations and standards, apart from the organisation's internal strategy. DevOps and security engineers are usually mandated to comply and follow certain frameworks. FIDO is a fast, scalable and extensible authentication framework that supports NIST 800-63, GDPR \cite{fido_whitepaper_Privacy_Principles}, PSD2\cite{fido_whitepaper_psd2_compliance}, KYC \cite{9045148} and PCI-DSS compliance. Especially in the case of the latest release of NIST 800-63 guidelines, where different levels of authenticator assurance are introduced, the Authenticator Assurance Levels (AAL), FIDO's stakeholders such as Yubico, identiv and queraltinc released new authenticators in order to reach AAL3 and receive FIPS 140-2 validation, which is the optimal in terms of security. Most of the FIDO authenticators are rated as AAL2 due to the fact that browsers do not support token binding to deliver Verifier Impersonation Resistance, as in Fig.~\ref{fig:FIDO_cert_and_levels}.

\subsection{Extensibility}
\label{subsec:extensibility}
FIDO protocol architecture is aiming towards a decentralised solution by bringing the authentication into the user's equipment, minimising the account recovery requests. Nevertheless, the FIDO server is responsible for verifying the result of the authenticator. For the purposes of identification and authorisation though, FIDO is usually combined and used with federated identity and authorisation protocols such as OpenID Connect and OAuth.
FIDO has been deployed with different themes that provide a linkage with a real identity. K-FIDO is an example of an integration widely adopted and accepted by end users to perform daily tasks. Several EU countries also suggest an integration with eIDAS while there was a research for the implementation of FIDO for the  Brazilian eGov Program\cite{8795121}. Due to the latest travel restrictions with COVID, even digital passports \cite{politis_covid_passports} or mobile driver's licence and car keys that use FIDO have been proposed \cite{google_car}. 
\par
These solutions though imply the co-existance with a third-party, which is not an optimal solution in terms of privacy. Therefore, the integration of Decentralised ID's seems to be the next step in the authentication as well as the identity and access management (IAM) ecosystem. The latest W3C specifications of WebAuthn dedicate a section to privacy issues that may occur. To this extend, IBM Security Verify solution  proposes a set of authenticator extensions for enhancing FIDO's authentication capabilities and allowing ID-less and passwordless authentication with QRCodes and FIDO, using Keycloak and/or RedHat SSO \cite{ibm_verify}. In \cite{8489316} a Decentralised ID (DID) system is proposed which is UAF compatible. A FIDO-based, privacy preserving solution is proposed in \cite{8931622} while \cite{s21082686} demonstrates how a solution as such could be used in the future smart grid.
\par
FIDO UAF although has not undergo a major update since 2017, is still applicable in a variety of applications and services, according to the survey. In particular, in \cite{8484268} an authentication and authorization flow for Wi-Fi enterprise networks, using FIDO UAF, is suggested. Nevertheless it is limited to mobile phones with Tursted Execution Environment (TEE) \cite{8484268} while both \cite{8166453} and \cite{CHIFOR2018740} propose FIDO UAF, IoT device support, leveraging of a smartphone. Samsung has offered an API \cite{broadcom_mag_and_samsungSDS} for FIDO support. Even though Samsung's Tizen brings FIDO UAF clients to devices such as mobiles and smartwatches \cite{samsung_tizen},\cite{samsung_tizen_native} it hasn't been implemented yet to support Smart TV's \cite{samsung_tizenFX}. 
\par
The future of machine-to-machine authentication is expected to be intrinsically linked to FIDO\cite{9319261}. Hints have been given though in UAF specifications which have introduced the concept of silent authenticators. Although there was a draft of FIDO IoT since 2020, the standard was not released released till March 2021 \cite{fdo}. The FIDO IoT specification covers both devices with IP and without IP capabilities that contain Rendezvous information and Ownership voucher. Although, the Fido Device Onboard (FDO) is designed for efficient, secure and resource aware onboarding of restricted IoT devices, it limits its applicability to device and network specific vertical applications, since it does not reuse the existing FIDO authenticators, excluding most of the current mobile devices used by home users. The conducted survey, clearly showcases that the FIDO protocol suite, includes a variety of authentication mechanism suitable for the rapidly evolving requirements of the Beyond 5G era, spanning heterogeneous vertical industries and satisfying stringent security requirements.

\subsection{Browser and OS-level support}
\label{subsec:browser_os_support}
Microsoft Azure is the most complete FIDO2 solution to this day. It has been successfully integrated in various use cases especially in the corporate ecosystem Citrix and Okta workspace, leverage of the Microsoft Azure Active Directory to offer passwordless authentication through FIDO2 \cite{citrix} \cite{okta}. Microsoft advises developers that want to support FIDO2 in their desktop applications to use either the \texttt{"Microsoft Authentication Library (MSAL)"} with the \texttt{"Windows Authentication Manager (WAM)"} or the WebView2 in an embedded browser \cite{microsoft_msal}. Microsoft's native iOS and Android support is under development.
\par
Apple offers WebAuthn API's since iOS \& iPadOS 14 and macOS Big Sur 1, announced in July 2020, focusing on the privacy aspect by adopting anonymous attestation. Apple's attestation format has been added in the latest WebAuthn API Level 2 specification \cite{w3c_webauthn_api}. In the Apple statement format the AAGUID is all zeros, to conceal user's privacy \cite{webkit}. In Safari, NFC, USB, and Lightning FIDO2-compliant security keys are available via \texttt{"SFSafariViewController"} and \texttt{"ASWebAuthenticationSession"}\cite{apple_webauthn_key_support}. Regarding UAF support for iOS \& iPadOS, apple has not relaeased any documentation up to this day. There are though commercial SDK's that developer's may use. 
\par
In Linux FIDO is provided by \texttt{"libfido2"} library, developed by Yubico and \texttt{"OpenSSH"} which offer U2F and CTAP2 support\cite{arch_linux_2ndfactor}. In addition, Yubico has developed a module for passwordless authentication via the \texttt{"Pluggable Authentication Module (PAM)"} framework for system-wide user authentication \cite{arch_linux_pam}. A FIDO package, based on \texttt{"libfido2"} library, is available in popular Linux distributions such as Arch Linux, CentOS, Fedora, Mageia, openSUSE, Solus as well as Debian and Ubuntu Linux \cite{pkg_fido_linux},\cite{debian_fido_linux}. Regarding commercial distributions, Red Hat Enterprise Linux offers UAF and U2F functionality via in RH-SSO and Keycloak \cite{RH_fido},\cite{ibm_verify} while Oracle Linux Yum Server added FIDO2 functionality in within 2021 \cite{oracle_yum_Server}. 
\par 
Android developers experienced a time advantage since it was the first mobile OS to release FIDO2 platform support for bound/internal authenticators on 2019. The Android API for FIDO client currently supports U2F and FIDO2 (WebAuthn client) \cite{Android_API_FIDO}, while CTAP2 support for external multi-factor authenticators is yet to be released. For CTAP2 external authenticator support, developers are still working with workarounds although it is speculated that updates will be released in 2021. 

\subsection{Implementation difficulties}
\label{subsec:implementation_difficulties}
\par
Although FIDO UAF was released prior to FIDO2/WebAuthn, there is no reference to UAF support by Android. Mobile application developers have been implementing UAF client through the \textit{com.uaf.fidoalliance} package whose tag in the .xml file is:  
\texttt{"org.fidoalliance.uaf.permissions.FIDO\char`_CLIENT"}. While there are specialised SDK's to enable the communication between the client application and the authenticator, these solutions are not open source. The performed reverse engineering confirmed that many services with certified FIDO UAF clients, do not use them with the latest version of their application. One possible explanation for the degradation of UAF could be the wider adaptation of FIDO2. To this extend, numerous services are shifting towards the support of FIDO2, which is the newest protocol. The reasons behind this wider acceptance of FIDO2 vary. Firstly, many services, such as government applications, do not offer specialised mobile applications therefore FIDO UAF is, by default, not an option. Secondly, the integration of FIDO UAF seems to be more challenging for developers, while FIDO2, due to its client support by all major web browsers, seems to be a more tangible option. Before the release of FIDO2, developers had to write a threefold code to cover  i) the communication of the firmware with the FIDO client application, ii) the communication of the latter with the client application as well as iii) the overall communication with the server. FIDO2 introduced a standards-based approach which is supported by the major browsers as well as some OS. To this extend, the work of developers has been limited to a twofold communication between i) the authenticator and the client application and ii) the latter with the server, which is supported by standardised API's. This level of abstraction allows less assumptions to be made by developers who are often non-security experts while it reassures frequent updates and patches as well as continuous support both from the developer's side through a wide developer's community and from the end-user's side since browsers provide support to older devices as well. This assumption is also backed up by the search performed on GitHub. UAF public repositories were limited whereas for WebAuthn the list of public projects was extended. 
\par
Studying Table~\ref{tab:github_repos} one may deduce that languages used mainly for back-end web programming (i.e., JavaScript, Java, Python, PHP, Ruby) are more popular for FIDO development. This finding applies to WebAuthn's search as well, in alignment with WebAuthn's scope that is the support of web applications running on FIDO-enabled clients which are the browsers. Languages that are essentially used for system programming, such as Rust and C, seem to be more popular in CTAP than WebAuthn, which is in-line with the protocol's description.
U2F search results contained both high and low level languages which is explained by the fact that U2F is not separated into two sub-protocols such as FIDO2 with WebAuthn and CTAP2.
For UAF, Java was the leading language, followed by other languages mainly used for web development. Here the fact that low-level programming languages were missing is particularly interesting. Even though Android does not support UAF in its API, Java is very popular while our findings demonstrate insufficient support by the open source community when it the communication with the authenticator, which is by default a low-level system procedure. Another interesting finding is that Swift, which is mainly used for iOS programming, was present in the WebAuthn and UAF queries. It was observed that, in absolute values, the number of projects between U2F and CTAP that use C or C++, which are considered as unsafe languages due to memory issues, was decreased. On the contrary, the use of Rust language was increased from version 1 to version 2 of the protocols, dictating a turn to safer programming languages \cite{264140}.
\par
A further drawback in the implementation of the FIDO protocols is the account recovery process. In order to overcome this issue, the FIDO Alliance recommends either enrolling more authenticators  or re-running identity proofing. Relying parties, apart from encouraging users to enroll more than one authenticators, should provide a reporting option in case of theft or damage while they should revoke the stolen or lost credentials \cite{fido_recvoery}.

\subsection{Other challenges for developers}
\label{subsec:other_challenges}
A recent study by Votipka et al. \cite{247694} enlists some important factors that influence secure programming from the developers' side. This work concludes that i) improving the API documentation, ii) developing vulnerability-finding tools, and iii) educating the developers, could provide effective solution towards more reliant and secure programming. Even though FIDO specifications are generally detailed enough, certain improvements would admittedly facilitate developers as proven in the FIDO dedicated forum for developers (see \footnote{\url{https://groups.google.com/a/fidoalliance.org/g/fido-dev/c/OznyLnSA4Z0/m/UnT7_isXAgAJ}}). 
\par
Additionally, the CTAP2 specifications have been updated in order to cover biometric authentication for external authenticators that desire an assurance level such as AAL3 per NIST, as described in section \ref{subsec:NIST_psd2_GDPR}. The biometric is used to verify the user and unlock the private key to sign the server challenge. Nevertheless, the updated and presumably more detailed CTAP version 2.2 has yet to be made accessible to the public; fact that poses an "unfair" advantage to the members of the Alliance. 
\par
In line with previous works, this study supports that solely a well-written API documentation is not enough to engage developers with the FIDO solutions. Vulnerabilities caused by implementation faults due to misunderstanding, API misuse or limited security background, should be mitigated. Developers need to have a basic knowledge of the protocols as well as where they may intervene and configure parts of the communication and what configurations should be chosen depending on the security and privacy needs of their organisation and the service provided. Tools which aim to help developers in elevating their knowledge of the protocol and finding bugs should be offered as well. Accordingly to this educational need from the developer's side, \cite{Grammatop} presents a tool that analyses the WebAuthn traffic of applications, aiming to to speed up the debugging process and provide a direct insight into the WebAuthn requests and responses.
\begin{table}[ht]
\centering
\caption{Number of public Github repositories per language for the FIDO protocols}. 
\label{tab:github_repos}
\resizebox{0.6\textwidth}{!}{%
\begin{tabular}{@{}l|cccccc@{}}
\toprule
\multirow{2}{*}{\textbf{Language}} &\multicolumn{6}{c}{\textbf{Protocol}} \\
 &U2F &UAF &WebAuthn &CTAP &FIDO &FIDO2\\
\midrule
JavaScript &\textbf{17} &0 &\textbf{81} &5 &\textbf{57} &\textbf{21}\\
TypeScript &0 &0 &33 &1 &2 &2 \\
PHP &6 &1 &32 &0 &13 &6 \\
Go &7 &1 &21 &0 &13 &5 \\
None &6 &3 &20 &4 &56 &18 \\
Java &7 &\textbf{8} &19 &\textbf{7} &36 &12 \\
HTML &1 &1 &18 &1 &10 &9 \\
Python &8 &1 &16 &3 &28 &13 \\
Ruby &5 &0 &10 &0 &10 &1 \\
C\# &3 &0 &9 &1 &6 &3 \\
Elixir &1 &0 &6 &0 &3 &0 \\
CSS &1 &0 &4 &0 &6 &4 \\
Rust &7 &0 &4 &2 &7 &5 \\
C &6 &0 &3 &1 &13 &6 \\
Kotlin &0 &1 &3 &1 &5 &7 \\
C++ &2 &0 &3 &0 &9 &3 \\
Vue &0 &0 &3 &0 &1 &0 \\
Swift &1 &1 &2 &0 &5 &0 \\
PLpgSQL &0 &0 &2 &0 &0 &0 \\
Perl &2 &0 &0 &0 &3 &0 \\
Lua &1 &0 &0 &0 &1 &0 \\ 
Dockerfile &1 &0 &0 &0 &2 &0 \\ 
XLST &1 &0 &0 &0 &1 &0 \\ 
Objective C &1 &0 &0 &0 &2 &0 \\ 
Clojure &1 &0 &1 &0 &1 &0 \\ 
Nim &1 &0 &0 &0 &1 &0 \\ 
Haskell &0 &0 &1 &0 &0 &1 \\ 
Tex &0 &0 &1 &0 &1 &2 \\ 
Scala &0 &0 &1 &0 &0 &0 \\ 
Jupyter Notebook &0 &0 &1 &0 &2 &0 \\ 
Pascal &0 &0 &0 &0 &1 &0 \\
Makefile &0 &0 &0 &0 &1 &2 \\
TSQL &0 &0 &0 &0 &1 &0 \\
Shell &0 &0 &0 &0 &1 &0 \\
Dart &0 &0 &1 &0 &1 &0 \\
Vala &0 &0 &0 &0 &1 &0 \\
\hline
\textbf{Total} &\textbf{86} &\textbf{17} &\textbf{295} &\textbf{26} &\textbf{331} &\textbf{135}\\
\bottomrule
\end{tabular}%
}
\end{table}

\par
Even though FIDO can be considered legacy authentication  protocol, it has yet to become the dominate authentication protocol, eliminating the use of passwords, due to the inability of the computer science community to convince users and stakeholders that personal biometric data will remain private and the sovereignty of their ID information will be ensured.

\subsection{Discovered Vulnerabilities and Mitigation Techniques}
\label{subsec:discover_vulnerabilities}
The FIDO Alliance advises developers to align with standards and programming best practices such as the ones documented by OWASP\cite{owasp} and acquire FIDO Certification to guarantee compliance with FIDO's security and privacy requirements \cite{fido_whitepaper_enterprise_servers}.
Google offers an educational environment, where developers may build and test their first WebAuthn app \cite{Google_build_first_webauthn}. The FIDO Alliance provides a tool for "conformance self‐validation testing" prior to the certification process in order to assess the conformance to the standards.  \cite{fido_conformance_for_cert}. An open version of the tool is accessible in GitHub \cite{fido_conf_github}. Google has published a CTAP2 testing tool \cite{Google_CTAP2testtool} while Google's Chrome browser allows developers to use the "WebAuthn Tab" for testing their infrastructure, leveraging of virtual authenticators \cite{Google_webauthn_tab}. StrongKey provides material for step-to-step integration in \cite{strong_key_tool}.The next step after the self-valitation is the "interoperability testing" which verifies that different a FIDO client may work with the FIDO server \cite{fido_interoperability_for_cert}.
Open versions of the tools can be found on GitHub \cite{fido_interoperability_github}. If these self-validation and interoperability testing are successfully completed, the certification procedure takes place. 
\par
Undetected weaknesses though might still exist. 
Theoretical vulnerabilities in UAF have been presented in \cite{hu2016security}, \cite{10.1007/978-3-319-67639-5_11} and \cite{fengformal} but only \cite{li2020authenticator} managed to demonstrate a feasible attack. Moreover, attacks related to hardware security and side channel in U2F have been also described throughout the years.
Transport protocols supported by FIDO (i.e., BLE) may also present vulnerabilities.
Risk engines that use machine learning (ML) could be used in order to mitigate threats as such, by producing a user profile.  


\par
TPM attacks are becoming more sophisticated, allowing information theft by exploiting physical vulnerabilities. Microsoft proposes a novel, hardware root of trust, implemented by the Pluton security processor which offers password-less authentication. One of the  major security problems solved by Pluton is keeping the system firmware up to date across the entire PC ecosystem. Vulnerabilities such as the one described in \cite{2017-ccs-nemec}, 
 which affect the security of RSA algorithm in TPM's, would be resolved by the Pluton. A technology such as the one proposed by Microsoft is pivotal for the enablement of IoT security as well. \cite{microsoft_7_Properties_Secured_Devices}, \cite{microsoft_pluton}. TEE is also gaining attention in recent years both for DevOps Architects and attackers. In \cite{s21020520} a thoughtful analysis of attacks thus measures to enhance TEE security are presented. The authors in \cite{li2020authenticator} concluded that the \textit{FacetID} and \textit{CallerID}, used by the UAF protocol to check the User Agent and the UAF Client application respectively, cannot guarantee integrity during run-time, therefore malicious attacks can be launched when TEE is employed as an authenticator. Proposed solutions include an architecture similar to the TrustZone-based Integrity Measurement Architecture (TIMA) to validate the integrity of the operating system, third-party verification or application reinforcement and code obfuscating technology and root detection mechanisms.




\section{Conclusion}
\label{sec:conclusion}
FIDO protocol has already managed to engage a wide audience of stakeholders and end-users nevertheless, more work needs to be done in order to be established as the authentication solution that replaces the unsafe username and password scheme. The level of penetration of FIDO2 and U2F mainly as second authentication methods is evident, as indicated from the survey of the USA and Asian markets. From our analysis we conclude that FIDO2 is easier for developers to implement due to the browser and OS support and the amount of public repositories. The integration of FIDO2 allowed developers to understand the basis and familiarize with the FIDO protocols. At the same time though, we support that the latest UAF v1.2 uses the previous UAF versions as well as the lessons learned from FIDO2 to propose strong authentication even for devices that cannot support browser, such as resource scarce next generation IoT deployments in factories of the future and connected/autonomous vehicles. Evidently, FIDO UAF is far from obsolete. On the contrary, this study surveyed and showcased that FIDO protocols, which extend from FIDO UAF to FDO, can satisfy the rapidly evolving user and system requirements, security constraints and time-to-market demands of the different applications and services of the emerging 6G and Industry 4.0/5.0 era.

\section*{Acknowledgment}
The project leading to this application has received funding from the European Union’s Horizon 2020 research and innovation programme under grant agreement No 101016608 (H2020-ICT-41-5GPPP-EVOLVED-5G), from the European Union's Horizon 2020 Stimulating innovation by means of cross-fertilisation of knowledge program under the Grant Agreement No 824015 (H2020-MSCA-RISE-2018-INCOGNITO), and the Greek state funded Operational Programme Competitiveness, Entrepreneurship and Innovation 2014-2020 (EPAnEK) under the Grant Agreements NetPHISH-$T1E\Delta K-05112$. We would like to thank Prof. Xenakis, who with his capacity as coordinator of the H2020-DS-2014-1-ReCred the first EU funded project aiming to implement FIDO, provided valuable insight for the completion of this work.




\bibliographystyle{ACM-Reference-Format}
\bibliography{sample-base}

\end{document}